\documentclass[aps,prl,twocolumn,groupedaddress]{revtex4}

\usepackage{amsmath}
\usepackage{graphicx}
\usepackage{multirow}

\begin{document}

\title{Diabatic ramping spectroscopy of many-body excited states for trapped-ion quantum simulators }
\author{Bryce Yoshimura$^1$, W. C. Campbell$^2$, and J. K. Freericks$^1$}
\affiliation{$^1$Department of Physics, Georgetown University, 37th and O st. NW, Washington DC, 20057, USA}

\affiliation{$^2$Department of Physics and Astronomy, University of California, Los Angeles, California 90095, USA}

\date{\today}

\begin{abstract}
Due to the experimental time constraints of state of the art quantum simulations with trapped ions, the direct preparation of the ground state by adiabatically ramping the field of a transverse field Ising model becomes more and more difficult as the number of particles increase. We propose a spectroscopy protocol that intentionally creates excitations through diabatic ramping of the transverse field and measures a low-noise observable as a function of time for a constant field to reveal the structure of the coherent dynamics of the resulting many-body states. To simulate the experimental data, noise from  counting statistics and decoherence error are added. Compressive sensing is then applied to Fourier transform the simulated data into the frequency domain and extract the the low-lying energy excitation spectrum. By using compressive sensing, the amount of data in time needed to extract this energy spectrum is sharply reduced making such experiments feasible with current technology.  

\end{abstract}

\maketitle

\section{I. INTRODUCTION}

Predicting the behavior of complex many-body quantum materials, such as frustrated magnets, can be an intractable problem on a digital computer~\cite{sachdev1999, diep2005, moessner2006}. Feynman proposed the use of a quantum-mechanical simulator to efficiently solve these problems~\cite{feynman1981}. One successful platform that can model spin systems are ion trap emulators~\cite{friedenauer2008, kim2009, kim2010, islam2011, britton2012, islam2013}. The success of these trapped-ion quantum simulators stems from their long coherence times, precise spin-state quantum control and high fidelty. These successes have been observed in linear Paul traps, which have successfully performed quantum simulations with as many as 18 ions in a one-dimensional linear crystal~\cite{senko2013}, and the Penning trap, which employs a two-dimensional crystal in a single-plane to trap $\sim$300 spins~\cite{britton2012}. Using either a linear Paul trap or the Penning trap, it has been demonstrated that a spin-dependent optical dipole force can be applied to the crystal of ions to realize a tunable Ising-type spin-spin coupling~\cite{kim2009}.

To prepare the ground state of the transverse Ising model in trapped-ion quantum simulators, the system of spins is started in the ground state of a strong transverse magnetic field. The transverse magnetic field is then slowly reduced to zero. If the transverse magnetic field is decreased adiabatically, then the system of spins will stay in the ground state and this technique for preparing complex ground states is called adiabatic state preparation~\cite{lloyd1996, farhi2000}. One of the experimental complications is that as the number of ions increases and the energy gap decreases, keeping the total experimental duration below the coherence time can result in diabatic transitions out of the ground state~\cite{islam2013}. This is true even for optimized rampings~\cite{richerme2013}. We propose a spectroscopy protocol to probe the low-lying energy spectra of the system of spins that takes advantage of the diabatic excitations at different transverse magnetic field strengths, which we call diabatic ramping spectroscopy. The diabatic ramping spectroscopy measurement is made by holding the transverse magnetic field constant to make a low-noise measurement after the system has been excited to a coherent superposition of ground and excited states. An alternative spectroscopic method has recently been carried out that actively modulates the field magnitude and looks for a system response to the modulation frequency~\cite{senko2013}.  This method has been shown to be very effective at zero transverse field and has been used to create interesting quantum superposition states in a 4-spin system.  The method we develop here is passive in the sense that the system response itself contains the frequency information, which we show can in general reveal many spectral lines at once without requiring a scan of the modulation frequency.  These two methods are therefore complimentary, and we show here that diabatic ramping spectroscopy extends its parallel state detection ability well into the finite field range

We explore the diabatic ramping spectroscopy by simulating data for trapped ions driven near the center of mass mode. This realizes an infinite-range transverse field Ising model, if we ignore all the other phonon modes. The general transverse field Ising model Hamiltonian for $N_{part.}$ particles is given by 
\begin{equation}
	\hat{\mathcal{H}}(t) = -\sum^{N_{part.}}_{i < j} J_{ij}\hat{S}^{(z)}_i \hat{S}^{(z)}_j + B^{(x)}(t)\sum^{N_{part.}}_i \hat{S}^{(x)}_i
	\label{eq:ComHam}
\end{equation}
where $\hat{S}^{(\alpha)}_i$ are the spin-1/2 operators in the $\alpha = x$, $y$, $z$ directions for the $i^{th}$ ion and we set $\hbar = 1$. The infinite-range transverse field Ising model follows when all spin-spin couplings are the same so that $J_{ij} = J_{0}/N_{part.}$. The spin operators satisfy the following commutation relations 
\begin{equation}
	\left[\hat{S}^{(\alpha)}_i, \hat{S}^{(\beta)}_j \right] = i \epsilon_{\alpha \beta \gamma} \hat{S}^{(\gamma)}_i \delta_{ij}, 
	\label{eq:spincommutation}
\end{equation}
where the Greek letters represent spatial directions, the Roman letters are the lattice sites and $\epsilon_{\alpha \beta \gamma}$ is the antisymmetric tensor.
The total spin operator, $\hat{S}_{tot}^{(\alpha)} =\sum_i \hat{S}_i^{(\alpha)}$, simplifies the infinite-range transverse field Ising model from Eq.~(\ref{eq:ComHam}) into
\begin{equation}
	\hat{\mathcal{H}}(t) = -\frac{J_0}{2}\left[\frac{\left(\hat{S}^{(z)}_{tot}\right)^2}{N_{part.}} - \frac{1}{4}\right] +B^{(x)}(t) \hat{S}^{(x)}_{tot}.
	\label{eq:hamiltonian}
\end{equation} 
Here, we study the ferromagnetic state of the Ising model with positive $J_0$.

The infinite-range transverse field Ising model corresponds exactly to a special case of the  Lipkin-Meshkov-Glick (LMG) model~\cite{lipkin1964}, when the model is written in the quasi-spin formalism. The general LMG model was introduced as an exactly solvable Hamiltonian for a many-body system (the example of many-body system considered is a finite system of nuclei) to compare to various techniques and formalisms. The LMG model has subsequently been studied numerically and analytically~\cite{lipkin1964, meshkov1964, glick1964, newman1977, gilmore1978}. By considering these analytic and numerical studies of the LMG model, the infinite-range transverse-field Ising model breaks up into submatrices, where each submatrix block is an eigenstate of the $S^2_{tot}$, and the submatrix with $S^2_{tot} = N_{part.}/2$ includes the ground state and the excited states of interest when $J_0 > 0$. Also within each submatrix, the eigenstates of the LMG model, in the quasi-spin formalism, split into two groups one being symmetric and the other antisymmetric under an interchange of the z component of spins. The splitting of the eigenstates into two groups in the LMG model is known as spin-reflection parity for the infinite-range transverse-field Ising model.    

More specifically, the Hamiltonian commutes with the total spin operator $\hat{S}^2_{tot}$, so the Hilbert space is reduced from $2^{N_{part.}}$ to $N_{part.}+1$ states for the ferromagnetic system, where the ground state has spin $S^{(z)} = N_{part.}/2$. The eigenstates also have spin-reflection parity, that is, under the partial inversion transformation $\hat{S}^{(x)}_{tot} \to \hat{S}^{(x)}_{tot}$, $\hat{S}^{(y)}_{tot} \to -\hat{S}^{(y)}_{tot}$, $\hat{S}^{(z)}_{tot} \to -\hat{S}^{(z)}_{tot}$ the Hamiltonian and the spin commutation relations remain the same. Due to the spin-reflection parity, the ground state is only coupled to eigenstates with the same spin-reflection parity. Eigenstates with the opposite spin-reflection parity become degenerate with the eigenstates with the same spin-reflection parity when $B^{(x)} \to 0$. 
\begin{figure}
	\includegraphics[scale=0.3]{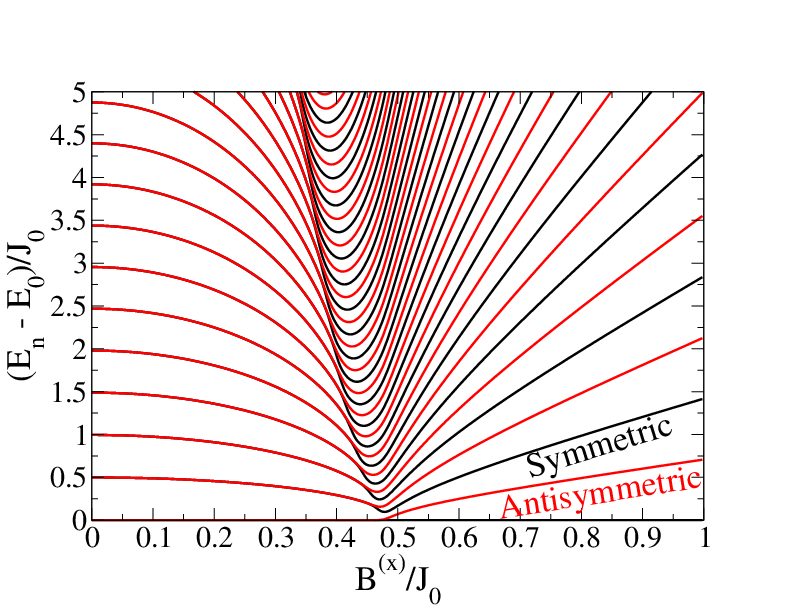}
	\caption{(Color online.) Example energy spectrum of the infinite-range transverse-field Ising model as a function of the transverse magnetic field for $N_{part.} = 400$ particles. The ground state can be excited to eigenstates with the same spin-reflection parity (black lines that alternate starting from the ground state). The eigenstates that have opposite spin-reflection parity (red lines that alternate in between the opposite parity lines) do not couple to the ground state, or any other opposite parity state. }
	\label{fig:energy_struct}
\end{figure}

Fig.~\ref{fig:energy_struct} shows the energy spectrum of the infinite-range transverse-field Ising model. Due to the avoided crossing of neighboring coupled eigenstates, a minimum energy gap occurs within each symmetry sector. The first minimum energy gap is between the ground state and the second excited eigenstate at a ``critical" transverse magnetic field strength that approaches $0.5J_0$ as the number of particles increases. The width of the first minimum energy gap is inversely proportional to the cube root of the number of particles, $E_2 -E_0 \propto N_{part.}^{-1/3}$~\cite{botet1983}. Following the first minimum energy gap, a second minimum energy gap occurs between the fourth and second excited eigenstates, and so on.

During an experiment where the transverse magnetic field is ramped to zero, excitations are primarily created when diabatically lowering the transverse magnetic field near the ``critical" transverse magnetic field strength $B^{(x)}(t) \approx 0.5J_0$. After excitation, we stop evolving the Hamiltonian at a specific time $t_{stop}$ and field $B^{(x)}(t_{stop})$ in order to perform an excited state spectroscopy measurement. The observable, $O_p(t)$, in the Heisenberg representation, evolves as a function of time with respect to $\hat{\mathcal{H}}(t_{stop})$, which is now a time independent Hamiltonian. The time evolution of the observable is given by the energy differences between eigenstates with the same spin parity as the ground state that have been diabatically excited, where $\hat{\mathcal{H}}(t_{stop})|m\rangle = E_m|m\rangle$ and
\begin{equation}
\hat{O}_p(t) = \sum_{mn} \langle m|\hat{O}_p |n\rangle \exp[-i(E_n - E_m)t].
\label{eq:oscillations}
\end{equation}
By analyzing this time dependence, one can extract the many-body energy differences.

The organization of the paper is as follows: In Sec. II, we outline the spectroscopy protocol and the methods used to simulate and process the data. In Sec. III, we provide representative numerical examples to illustrate how the energy spectra can be extracted by using signal processing. In Sec. IV, we provide our conclusions.  

\section{II. THEORETICAL FORMULATION}

\subsection{A. Spectroscopy protocol}
\begin{figure}
	\centering
	\begin{tabular}{c c c}
		\includegraphics[scale=0.27]{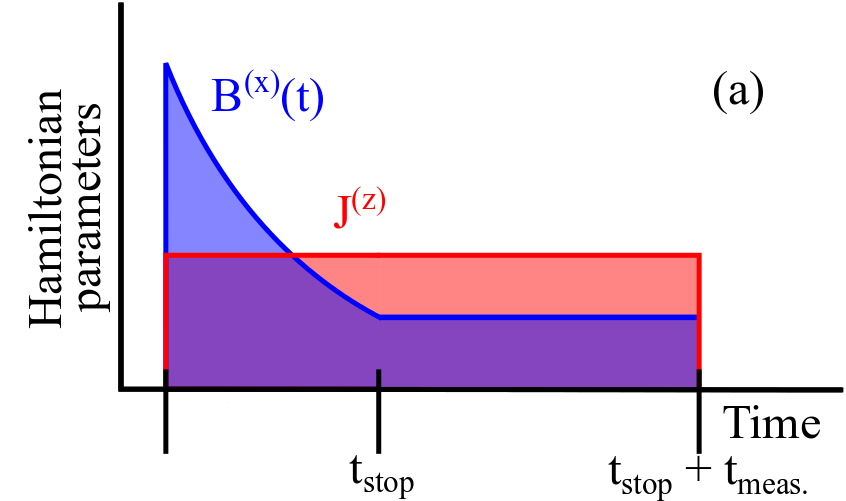}\\
		\includegraphics[scale=0.27]{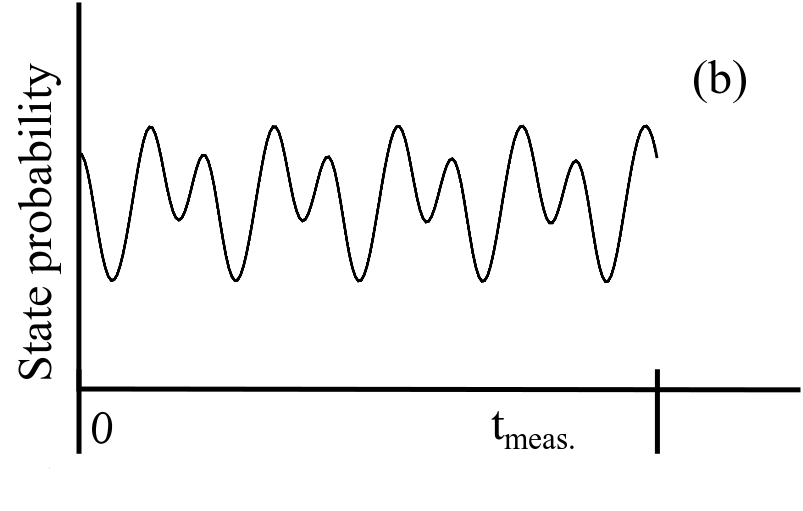}\\
		\includegraphics[scale=0.27]{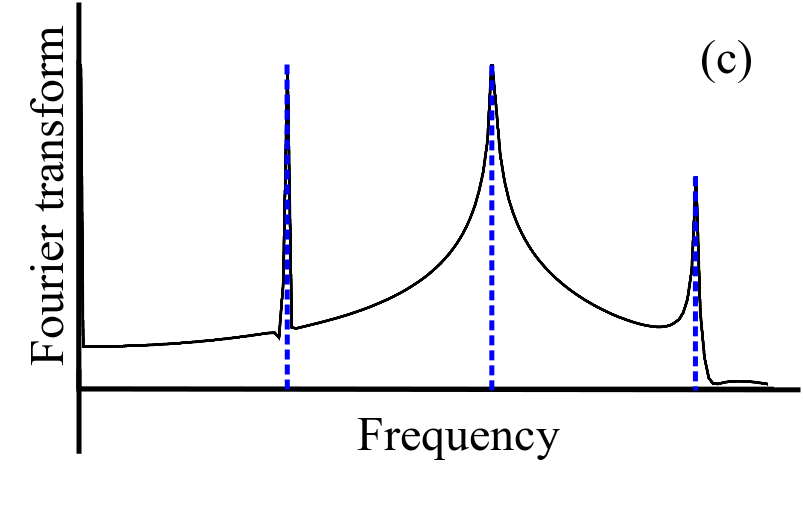}
		\end{tabular}
	\caption{(Color online.) Schematic diagram of the spectroscopy protocol. (a) The transverse magnetic field as function of time is diabatically ramped down to a chosen value, $B^{(x)}(t_{stop})$. $B^{(x)}(t_{stop})$ is then held for a time interval of $t_{meas.}$. (b) A low-noise observable measured during the interval $t_{meas.}$. (c) The low-noise observable as a function of $t_{meas.}$ is Fourier transformed into the frequency domain to determine the energy differences (solid line). After applying signal processing to the low-noise observable as a function of $t_{meas.}$ the energy differences can be determined more accurately (dashed lines).}
	\label{fig:cartoon}
\end{figure}

The energy spectra of the infinite-range transverse-field Ising model can be measured by creating excitations in the quantum simulation. The diabatic excitations depend on the rate at which the transverse magnetic field is ramped, and on the size of the minimum energy gap between the ground state and the first coupled excited state. The spectroscopy protocol is as follows (and depicted schematically in Fig.~\ref{fig:cartoon}):
\begin{enumerate}
	\item Diabatically decrease the magnetic field with time constant $\tau_{ramp}$ starting from a large polarizing field $B_0$ in the $x$ direction
	\begin{equation}
		B^{(x)}(t) = B_0 \mathrm{e}^{-t/\tau_{ramp}},
		\label{eq:magnetic}
	\end{equation}	
	evolving the quantum state and creating excitations, as shown in Fig.~\ref{fig:cartoon}(a).
	\item Decrease the magnetic field until the desired value is reached at $t = t_{stop}$ and then hold the field constant for a fixed time interval $t_{meas.}$, as shown in Fig.~\ref{fig:cartoon}(a).
	\item Measure a low-noise observable of interest at each $t_{meas.}$.
	\item For each new $t_{meas.}$, steps 1-3 are repeated for the necessary number of time steps to perform signal processing, as depicted in Fig.~\ref{fig:cartoon}(b). 
	\item Signal process (Fourier transform) the oscillations of the low-noise observable as a function of time to determine the energy differences, as shown in Fig.~\ref{fig:cartoon}(c).
	\item Repeat the protocol at different stopping values of the transverse magnetic field to map the energy spectra versus transverse magnetic field.
\end{enumerate}
The frequencies of these excitations are extracted from the measured signal as a function of time by Fourier transforming into the frequency domain. In the frequency domain, the exact signal will have peaks at the frequencies of the excitation energies; they will be broadened if the measurement has decoherence, noise, or a finite time evolution window. 

\subsection{B. Time evolution}
In order to evaluate the time dependence of the observable, we must evaluate the time evolution with respect to the time-dependent Hamiltonian. We do this with the evolution operator, which satisfies
\begin{equation}
	i\frac{\partial}{\partial t} \hat{U}(t, t_0) = \hat{\mathcal{H}}(t) \hat{U}(t,t_0)
\end{equation}
and $\hat{U}(t_0, t_0) = 1$. Since the total spin operators have the same commutation relations as in Eq.~(\ref{eq:spincommutation}), the Hamiltonian does not commute with itself at different times (during the ramp, $0 < t < t_{stop}$) 
\begin{equation}
	\left[\hat{\mathcal{H}}(t),\hat{\mathcal{H}}(t') \right] \neq 0,
	\label{eq:Hamcommute} 
\end{equation}
where $t \neq t'$. As a result of Eq.~(\ref{eq:Hamcommute}), the evolution operator must be calculated as a time-ordered product. We apply the evolution operator $\hat{U}(t, t_0) = \mathcal{T}_t \exp\left[-i \int_{t_0}^t \mathrm{d}t'\hat{\mathcal{H}}(t')\right]$ acting on the initial quantum state $|\psi(t_0)\rangle$ to determine the time evolution
\begin{equation}
	|\psi(t)\rangle = \hat{U}(t, t_0) |\psi(t_0)\rangle.
\end{equation}
The simplest way to evaluate the evolution operator is via a Trotter product equation~\cite{trotter1959}
\begin{equation}
	\begin{split}
		\hat{U}_{Trotter}(t, t_0) = &\hat{U}_{mid}(t, t- \delta t) \hat{U}_{mid}(t - \delta t, t-2\delta t) \dots\\ &\times \hat{U}_{mid}(t_0 + \delta t, t_0).
	\end{split}
\end{equation}
that is evaluated with a midpoint integration rule
\begin{equation}
	\hat{U}_{mid}(t + \delta t, t) = \exp \left[-i \delta t \mathcal{H}(t + \delta t/2) \right] 
\end{equation} 
for each factor in the product.

The error of the midpoint integration approximation scales as $(\delta t)^2$,  as can be seen by recombining products of exponentials into exponentials of the sums of the arguments. For example, the product
\begin{equation}
	\exp\left[-i\delta t \hat{\mathcal{H}}(\bar{t} +\delta t/2)\right]\exp\left[-i\delta t \hat{\mathcal{H}}(\bar{t} - \delta t/2)\right]
\end{equation}
is recombined using the  Baker-Campbell-Hausdorff (BCH) theorem~\cite{campbell1897, baker1902, hausdorff1906} 
\begin{equation}
	\begin{split}
		\mathrm{e}^X\mathrm{e}^Y = &\exp\left[ X+ Y + \frac{1}{2}\left[X, Y\right] \right. \\
 		&\left. + \frac{1}{12} \left(\left[X, \left[X, Y\right]\right] - \left[Y, \left[X, Y\right]\right]\right) + \cdots \right],
	\end{split}
	\label{eq:BCH}
\end{equation}
with $X = -i\delta t \hat{\mathcal{H}}(\bar{t} - \delta t/2)$ and $Y = -i\delta t \hat{\mathcal{H}}(\bar{t} + \delta t/2)$. The commutator term is proportional to $(\delta t)^2$ which is called the Trotter error. To reduce the error of the midpoint integration approximation, we use the commutator-free exponential time (CFET)~\cite{alverman2011, alverman2012} approach that utilizes a product of exponentials to determine each Trotter factor. The essential idea of the CFET procedure is to construct the Trotter factor such that when the product of Trotter factors are combined using the BCH formula, in Eq.~(\ref{eq:BCH}), the resulting expression is equal to a high-order truncated Magnus expansion~\cite{magnus1954} of the evolution operator
\begin{equation}
	\begin{split}
		\hat{U}(t,t_0 ) &= \exp \left[-i \left( \int_{t_0}^t \mathrm{d}t_1 \, \hat{\mathcal{H}}(t_1) \right.\right. \\
		&+ \left. \left. \frac{1}{2}\int_{t_0}^t\mathrm{d}t_1\int_{t_0}^{t_1} \mathrm{d}t_2 [\hat{\mathcal{H}}(t_1), \hat{\mathcal{H}}	(t_2)]+\cdots\right) \right]
	\end{split}
\end{equation} 
with as high an order expansion as possible. We use the optimized fourth-ordered CFET procedure that has an error of order $(\delta t)^5$. 

The optimized fourth-ordered CFET, $\hat{U}_{CFET}(t+\delta t, t)$, approximates the evolution operator with piecewise propagation for Hamiltonians of the form $\hat{\mathcal{H}}^{(z)} + B^{(x)}(t)\hat{\mathcal{H}}^{(x)}$ with $\hat{\mathcal{H}}^{(z)} = -J_0\left(\left(\hat{S}^{(z)}_{tot}\right)^2/2N_{part.} - 1/8 \right)$ and $\hat{\mathcal{H}}^{(x)} = \hat{S}^{(x)}_{tot}$ as follows:
\begin{equation}
	\begin{split}
		\hat{U}_{CFET}&(t+\delta t, t) =\\ &\exp\left[\delta t_1 \left(-\frac{J_0}{2}\left(\frac{\left(\hat{S}^{(z)}_{tot}\right)^2}{N_{part.}} -\frac{1}{4}\right) + 		\mathit{b}_1\hat{S}^{(x)}_{tot}\right)\right] \\
		 \times &\exp\left[\delta t_2 \left(-\frac{J_0}{2}\left(\frac{\left(\hat{S}^{(z)}_{tot}\right)^2}{N_{part.}} -\frac{1}{4}\right) + \mathit{b}_2\hat{S}^{(x)}_{tot}\right)\right] \\
 		\times &\exp\left[\delta t_1 \left(-\frac{J_0}{2}\left(\frac{\left(\hat{S}^{(z)}_{tot}\right)^2}{N_{part.}} -\frac{1}{4}\right) + \mathit{b}_3\hat{S}^{(x)}_{tot}\right)\right],
	\end{split}
	\label{eq:CFET}
\end{equation}
with time steps  
\begin{equation}
	\delta t_1 = \frac{11}{40}\delta t, \quad \delta t_2 = \frac{9}{20}\delta t.
\end{equation}
The magnetic field is evaluated at three different times in the interval of size $\delta t$ (with $x_i \in [0,1]$) 
\begin{equation}
	x_1 = \frac{1}{2} - \sqrt{\frac{3}{20}}, \quad x_2 = \frac{1}{2}, \quad x_3 = \frac{1}{2} + \sqrt{\frac{3}{20}}.
\end{equation}
The $b_1,\, b_2,\, b_3$ coefficients are calculated from
\begin{equation}
	\begin{pmatrix}
		b_1 \\
		b_2 \\
		b_3
	\end{pmatrix}
	=
	\begin{pmatrix}
		h_1 & h_2 & h_3 \\
		h_4 & h_5 & h_4\\
		h_3 & h_2 & h_1
	\end{pmatrix}
	\begin{pmatrix}
		B^{(x)}(t + x_1 \delta t) \\
		B^{(x)}(t + x_2 \delta t) \\
		B^{(x)}(t + x_3 \delta t)
	\end{pmatrix}
\end{equation}
where the elements of the matrix are
\begin{equation}
	\begin{split}
		h_1 = \frac{37}{66} -\frac{400}{957}&\sqrt{\frac{5}{3}}, \quad h_2 = -\frac{4}{33}, \quad h_3 =  \frac{37}{66} +\frac{400}{957}\sqrt{\frac{5}{3}}, \\
		& h_4 = - \frac{11}{162}, \quad h_5 = \frac{92}{81}.
	\end{split}
\end{equation} 
The $h$ coefficients were determined by taking the set of $b$ factors in Eq.~(\ref{eq:CFET}), combining them using the BCH formula, and then setting them equal to the truncated Magnus expansion of the evolution operator over the $\delta t$ time interval. This yields the reduced numerical error without needing to evaluate any commutators. Details can be found in Refs.~\citep{alverman2011, alverman2012}

\begin{figure}[t!]
	\centering
	\begin{tabular}{c c}
		\includegraphics[scale=0.09]{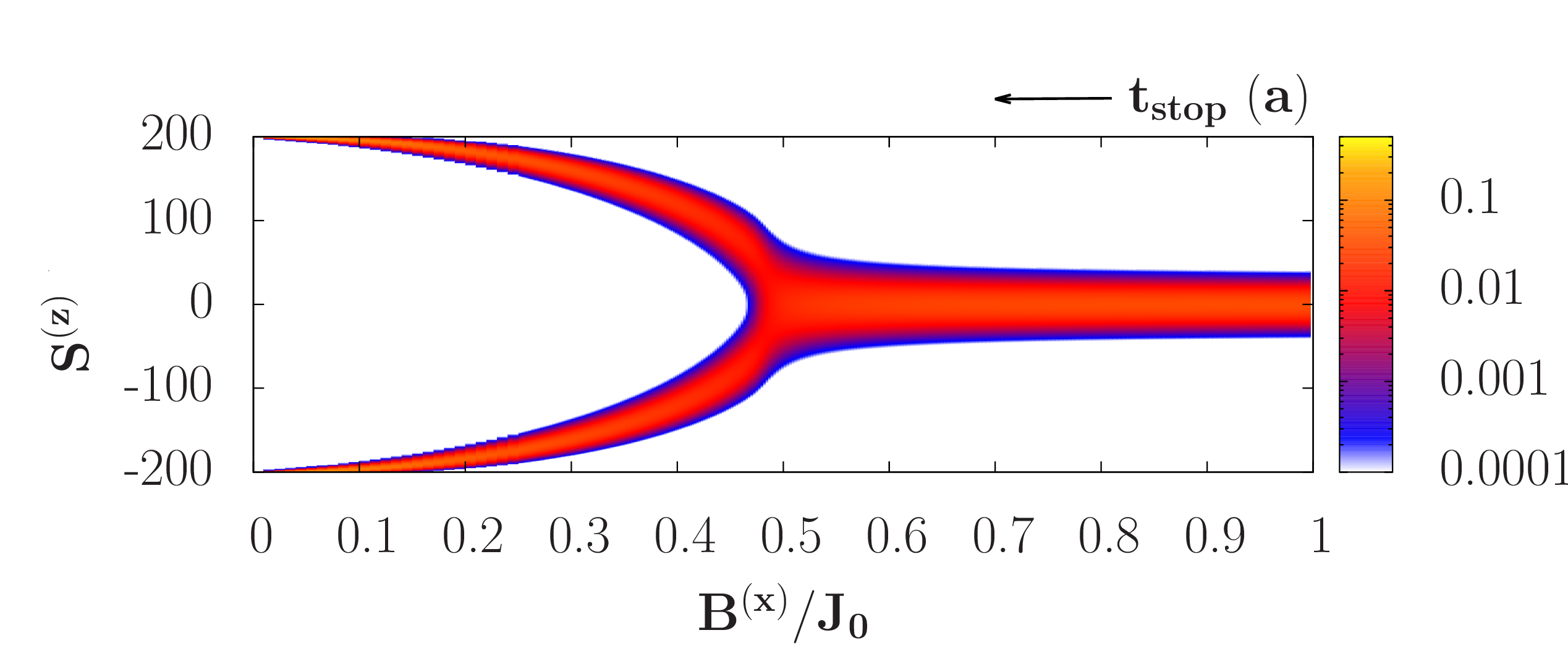}\\
		\includegraphics[scale=0.09]{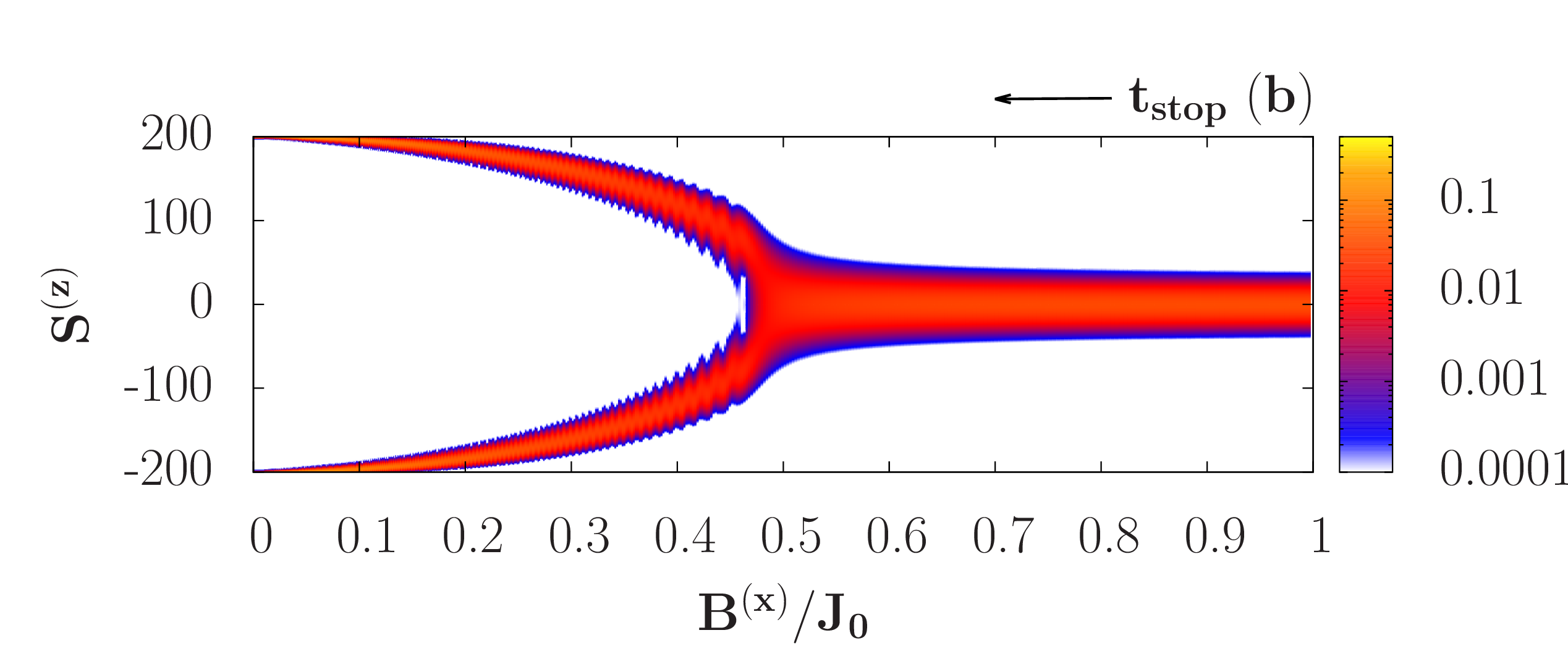}\\
		\end{tabular}
	\caption{(Color online.) Probabilities of the product states as a function of the transverse magnetic field, $B^{(x)}$, in the (a) adiabatic limit and (b) the numerically evolved diabatic case with the ramping rate of $\tau_{ramp} J_0= 2$. In both cases the highest probable $S_{tot}^{(z)}$ state starts at $S_{tot}^{(z)} = 0$ and moves towards $S_{tot}^{(z)} = \pm 200$ as the $B^{(x)}$ approaches 0. While the adiabatic limit shows the transition from the low to high $S_{tot}^{(z)}$ is smooth, the numerical results have ripples after the minimum energy gap due to the diabaticity of the time evolution.}
	\label{fig:probability}
\end{figure}

\subsection{C. Decoherence and noise}
We choose to measure the time dependence of the occupancy of the highest probable product state in the $S^{(z)}$ basis at time $t_{stop}$ for the observable as a function of time, $p_{exact}(t)$, during the fixed time interval $t_{meas.}$. The highest probable state stays at $S_{tot}^{(z)} = 0$ as $B^{(x)}(t)$ approaches the minimum energy gap. After the minimum energy gap, the highest probable state symmetrically moves toward $S_{tot}^{(z)} = \pm N_{part.}/2$, as shown in Fig.~\ref{fig:probability}.

\begin{figure}
	\includegraphics[scale=0.3]{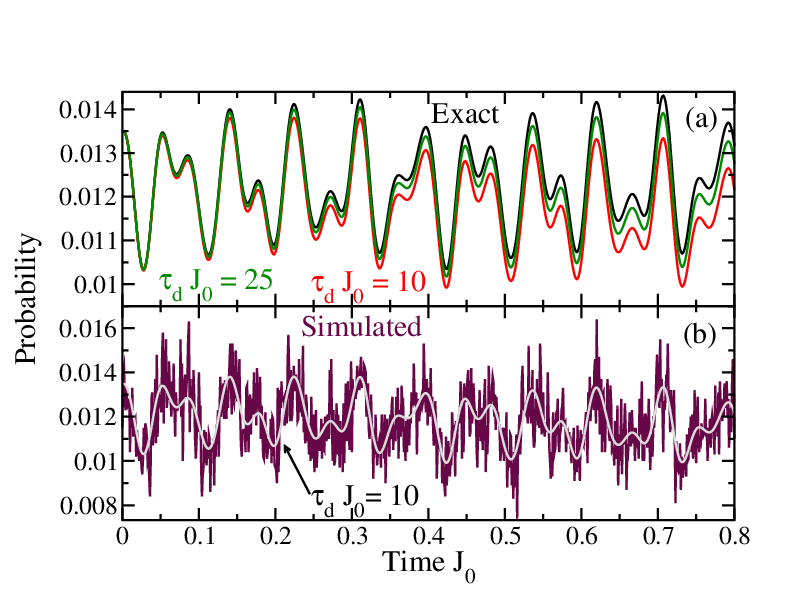}
	\caption{(Color online.) (a) Comparison of the exact signal (black) to a signal with decoherence added in via Eq.~(\ref{eq:decoherence}) ( red $\tau = 10/J_0$; green $\tau = 25/J_0$ ). (b) Counting statistics is added in by choosing random integers from the Poisson distribution with a mean value of $N_{meas.} p_{signal}(t)$, for $N_{meas.} = 10,000$ (violet).}
	\label{fig:noise}
\end{figure}

To simulate experimental data, we need to introduce typical errors. The two sources of error that we introduce are decoherence of the signal and counting statistics noise. The decoherence is modeled by a simple exponential decay of the exact signal
\begin{equation}
	p_{signal}(t) = p_{exact}(t) \mathrm{e}^{-\frac{t}{\tau_d}},
	\label{eq:decoherence}
\end{equation}
where $\tau_d$ is the decoherence time. We used a decoherence times of $\tau_d J_0= 25$ and $10$ in Fig.~\ref{fig:noise}(a). The counting statistics noise is added by randomly choosing integers from a Poisson distribution to represent the number of counts for the low-noise observable of interest, $p_{simulated}(t)$, at time $t$, as seen in Fig.~\ref{fig:noise}(b). The Poisson distribution is
\begin{equation}
	Pois(x|\lambda) = \frac{\lambda^x}{x!}\mathrm{e}^{-\lambda}
\end{equation}
where $\lambda$ is the mean of the distribution and $x = p_{simulated}(t)$ is the actual occurrence of an event. The mean value is $\lambda = N_{meas.} p_{signal}$, where $N_{meas.}$ is the total number measurements made at time $t$. To randomly choose an integer $x$ from a Poisson distribution, a random number $u_i$ is chosen from a uniform distribution (with $u_i \in [0, 1]$) $x$ number of times and the $u_i$'s are multiplied together~\cite{knuth1998}. When the product of $u_i's$ is less than $\mathrm{e}^{-\lambda}$, $p_{simulated}(t)$ is set equal to $x$. 
\begin{equation}
\prod^{x = p_{simulated}(t)}_{i=1} u_i < \mathrm{e}^{-\lambda}
\label{eq:poisson}
\end{equation}
The number of total measurements, $N_{meas.}$, at each time step was determined when the signal-to-noise ratio (SNR) of the initial probability at $t_{stop}$ is larger than $1$
\begin{equation}
	SNR \approx \sqrt{N_{meas.} p_{signal}(0)} > 1.
\end{equation} 
 
\subsection{D. Signal processing}
The signal processing of the oscillations of the low-noise observable is usually measured at equally spaced time steps, $t_n = n\delta t$, for a fixed time interval $t_{meas.} = (N_{step} - 1) \delta t$ and is transformed into the frequency domain by applying the discrete Fourier transform 
\begin{equation}
	 	P_{simulated}(f_k) = \frac{1}{\sqrt{N_{step}}}\sum_{n = 0}^{N_{step}-1} p_{simulated}(t_n)\mathcal{F}_{k,n},
\end{equation}
where $f_k = k/(N_{step}\delta t)$ for $|k| < N_{step}/2$ and $\mathcal{F}_{k,n}$ is
\begin{equation}
	\mathcal{F}_{k,n} = \mathrm{e}^{2i\pi k n/N_{step}}.
\end{equation}
The signal in the frequency domain can then be transformed back to the time domain by the inverse discrete Fourier transform
\begin{equation}
 		p_{simulated}(t_n) = \frac{1}{\sqrt{N_{step}}}\sum_{k = -\frac{N_{step}}{2}}^{N_{step}/2} P_{simulated}(f_k)\mathcal{F}^{-1}_{n,k}.
\end{equation}
Due to the noise of the oscillations and the characteristics of the discrete Fourier transform (as given by the Nyquist-Shannon analysis~\cite{nyquist1928, shannon1949}), the number of measurements taken of the oscillations needs to be large to get a good estimate of the energy differences (or frequency of the oscillations). The number of measurements can be significantly reduced when the signal processing algorithm called compressive sensing~\cite{donoh2006} is used, as we discuss below.

\subsubsection{Discrete Fourier transform}
\begin{figure}[t!]
	\centering
	\begin{tabular}{c}
		\includegraphics[scale=0.2]{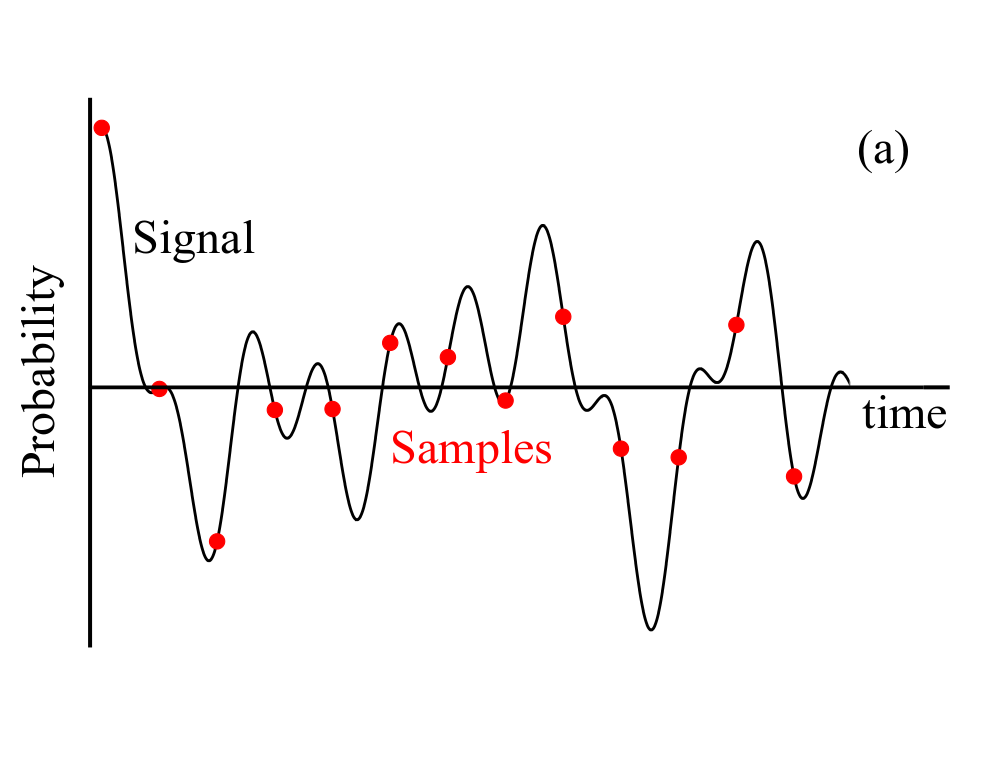}\\
		\includegraphics[scale=0.2]{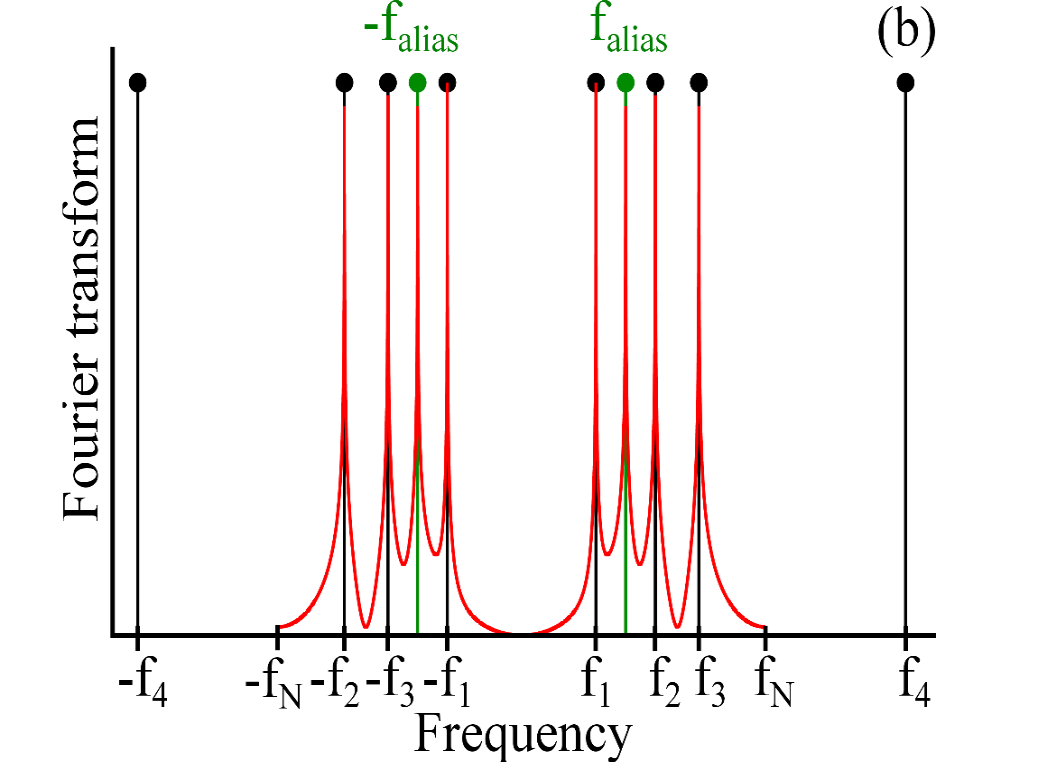}\\
		\includegraphics[scale=0.2]{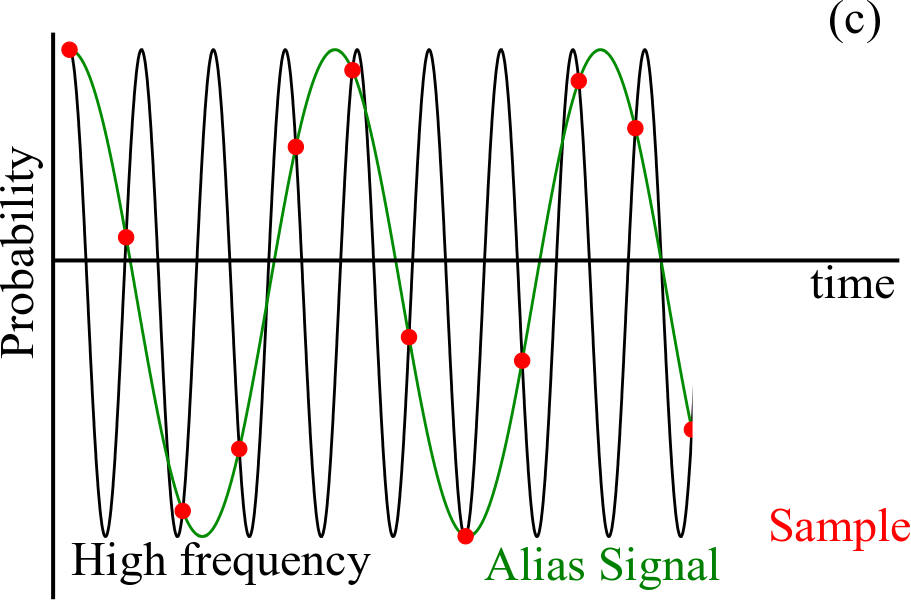}
		\end{tabular}
	\caption{(Color online.) (a) Signal (black line) that is made from four sine waves with their respective frequencies and sampled (red dots) at a frequency $1/\delta t$. The first three frequencies $f_1$, $f_2$, $f_3 < f_N$ and the fourth frequency, $f_4$, are greater than $f_N$. (b) The samples are Fourier transformed and $f_1$, $f_2$, $f_3$ (black line) can be identified from the signal as a function of time (red line). However, the fourth frequency appears in the signal as a function of time with an aliased frequency, as defined in Eq.~(\ref{eq:aliascondition}). (c) Comparing the sine wave of the high frequency, $f_4$, (black line) and the alias frequency, $f_{alias}$ (green line), the two are nearly indistinguishable with respect to the sampled points, (red dots) that were used on the original signal in panel (a).}
	\label{fig:alias}
\end{figure}

It is well known that $N_{step}$ equally spaced time steps of width $\delta t$ can determine the Fourier transform accurately for frequencies less than the Nyquist frequency~\cite{nyquist1928}, $f_N$,
\begin{equation}
	f_N = \frac{1}{2 \delta t}.
\end{equation}
This limit on the range of frequencies comes from the Nyquist-Shannon sampling theorem. Shannon proved that if the Fourier transform is nonzero for a finite frequency range, $|f| < f_N$, then the Fourier transform can be accurately determined with a sampling time that satisfies $\delta t = 1/(2f_N)$~\cite{shannon1949}. If the Fourier transform of the signal has frequencies, $f_{H,i}$, that are higher than the Nyquist frequency, $f_N$, then spurious data is generated due to a phenomenon called aliasing. Aliasing is when the frequencies that are higher than the Nyquist frequency are mapped into the range of frequencies that is less than the Nyquist frequency via 
\begin{equation}
	f_N > f_{H,i} - 2nf_N = f_{alias,i}, 
	\label{eq:aliascondition}
\end{equation}
where $n$ is an integer that satisfies the inequality, as illustrated in Fig.~\ref{fig:alias}(a-b). Eq.~(\ref{eq:aliascondition}) is determined by comparing the equally spaced time samples of a sine wave with a high frequency, $f_H$ and the aliased frequency, $f_{alias}$ which become nearly indistinguishable when the time steps are $\delta t < 1/(2f_H)$, as demonstrated in Fig.~\ref{fig:alias}(c).

\begin{figure}
\centering
	\includegraphics[scale=0.3]{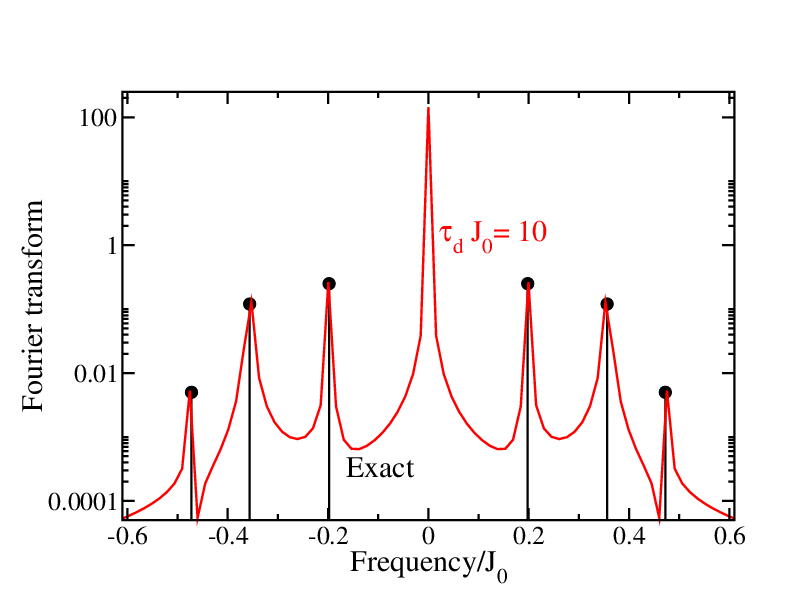}
	\caption{(Color online.) Fourier transformation of the probability as a function of time in Fig.~\ref{fig:noise}(a) with a decoherence time of $\tau_d J_0 = 10$ in the frequency domain. The probability was measured at $N_{step} = 2048$ equally spaced time steps. The delta function peaks of the signal have been broadened (red line) with respect to the Fourier transform without decoherence (black lines). }
	\label{fig:ftdecoherence}
\end{figure}

\begin{figure}[h!]
	\centering
	\begin{tabular}{c}
		\includegraphics[scale=0.3]{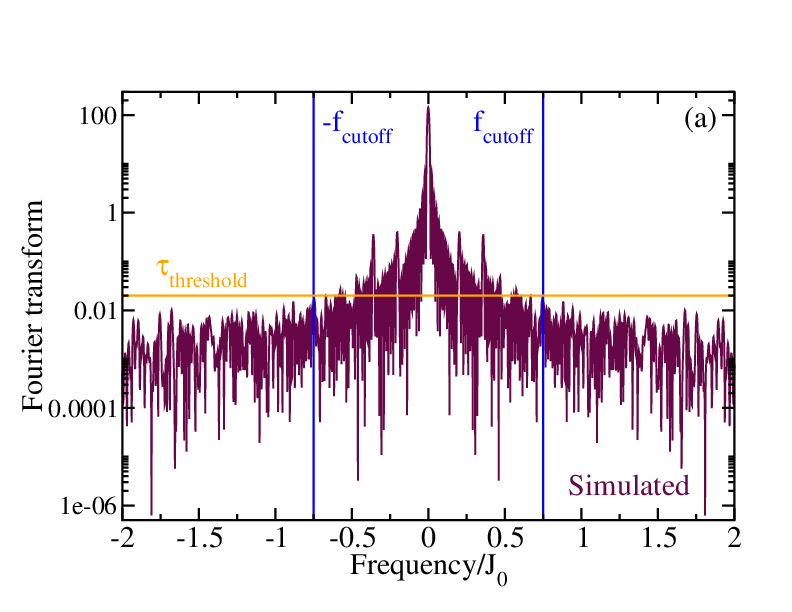}\\
		\includegraphics[scale=0.3]{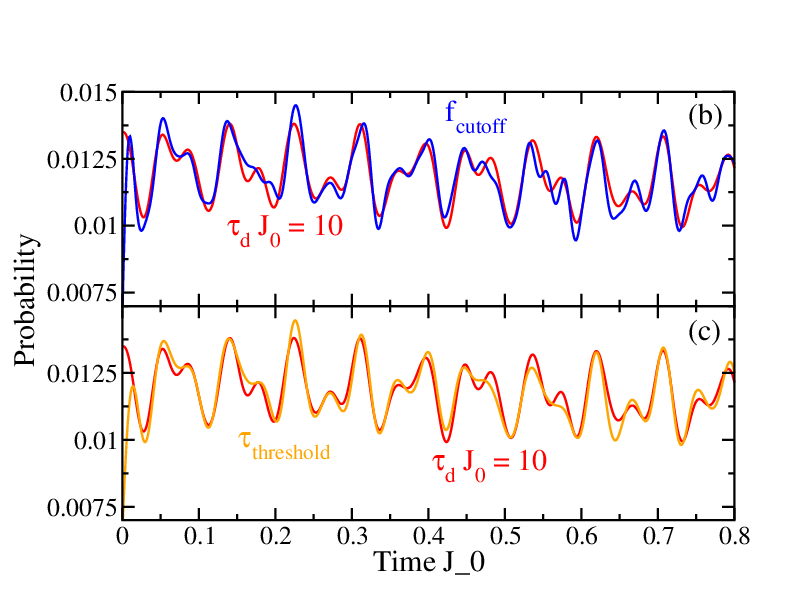}\\
		\end{tabular}
	\caption{(Color online.) (a) Fourier transform of the observable as a function of time in Fig.~\ref{fig:noise}(b) into the frequency domain, (violet line). There are two filters that can be applied in frequency domain: a low-pass filter and a thresholding filter. The low-pass filter sets all the frequencies $f > |f_{cutoff}|$ equal to 0 (where the blue lines represent the $f_{cutoff}$ applied in panel b), and the thresholding filter sets all the frequencies that have Fourier series coefficients lower than the threshold, $\tau_{threshold}$, equal to 0 (the amplitude of $\tau_{threshold}$ used in panel c is shown in yellow). (b) Low-pass filtering the signal from panel a and Fourier transforming the resulting signal to the time domain (blue), the counting statistics noise is reduced when comparing the resulting signal of Fourier transforming the low-pass filtered signal to the noiseless observable as a function of time with decoherence (red). (c) The resulting signal (yellow) of Fourier transforming the thresholding filtered observable as a function of frequency nearly lies on top of the noiseless observable as a function of time with decoherence (red).}
	\label{fig:compression}
\end{figure}

The effects of introducing decoherence and noise to the observable as a function of time, $p_{simulated}(t)$, produces errors in the Fourier transformation to the frequency domain. The decoherence error broadens the delta function peaks at the frequencies of the excitation energies in the frequency domain, as shown in Fig.~\ref{fig:ftdecoherence}. Once decoherence has contaminated the observable as a function of time not much can be done to reduce the effects of the broadening, unless there is only one decoherence time and it is known or can be fit. 
The counting statistics noise added to the observable as a function time is analogous to adding a linear superposition of nearly equally weighted sine waves that oscillate at a continuum of high frequencies to the signal, as depicted in Fig.~\ref{fig:compression}(a). When the observable as a function of time with the counting statistics noise is Fourier transformed to the frequency domain, the linear superposition of sine waves that oscillate at a continuum of high frequencies transform into a noise floor, since the sine waves have similar weight in the time domain. The noise floor then obscures the delta function peaks at the frequencies of the excitation energies with amplitudes below the noise floor. The simplest way to reduce the effects of the counting statistics noise on the observable in either the frequency or time domain is to increase the number of measurements, $N_{meas.}$ taken at each time step. To increase the amplitude of the delta function peaks above the noise floor, the length of the time interval, $t_{meas.}$, must be increased. However, when $t_{meas.} \gg \tau_{d}$ the observable as a function of time becomes nearly $0$.

If there is knowledge of the observable in the frequency domain that can restrict the frequencies, then the Fourier transform can be used to reduce the counting statistics noise that is present in the observable as a function of time. Once the observable is Fourier transformed to the frequency domain, two filters can be applied based on the knowledge of the observable in the frequency domain, as shown in Fig.~\ref{fig:compression}(a). If the observable in the frequency domain is known to have a delta function peak for a range of frequencies, then a low-pass filter is applied for a range of frequencies, $[-f_{cutoff}, f_{cutoff}]$, and all frequencies outside of this range are set to 0. Alternatively, if the noise floor can be estimated, then a thresholding filter is applied so that frequencies with a Fourier transform coefficient below a certain amplitude, $\tau_{threshold}$, are set to 0. The filtered observable as a function of frequency is inverse Fourier transformed into the time domain with a significant reduction of the counting statistics noise, as shown in Fig.~\ref{fig:compression}(b-c).      

\subsubsection{Compressive sensing}
However, when the Fourier series has weights only at $s$ discrete frequencies, the low-noise observable in the frequency domain has $s$ number of nonzero elements. To solve for the $s$ nonzero elements far fewer data should be needed. There have been significant advancements in signal processing to decrease the number of measurements determined from the sampling theorem when a signal has $s$ nonzero elements, called $s$-sparse, in a basis. This signal processing is called compressive sensing and the number of measurements, $M_{step}$ is limited by
\begin{equation}
	M_{step} \gtrsim s Log(N_{step}).
	\label{eq:condition}
\end{equation}
Although $M_{step}$ is greater than $s$, there is still a huge decrease in the running time for an experiment. Suppose we restrict the frequency interval to $[-f_c, f_c]$ and perform $N_{step.} = 10,000$ measurements in time using conventional Fourier transform techniques, then if the signal is known to have only three discrete frequencies, compressive sensing requires only $M_{step} = 200$ time steps for equivalent accuracy. For $N_{meas.} = 5000$ at a given time step, being taken at a rate of $100$ measurements per second, we would require about $3$ hours of time to generate the data for one magnetic field using compressive sensing. If the signal processing was done by the fast Fourier transform algorithm, it would take $139$ hours to collect the data at one magnetic field, using the same number of measurements at a given time step taken at the same measurement rate as before. Hence, compressive sensing makes this type of experiment feasible with current experimental setups.

\begin{figure*}[t!]
\centering
\includegraphics[scale=0.4]{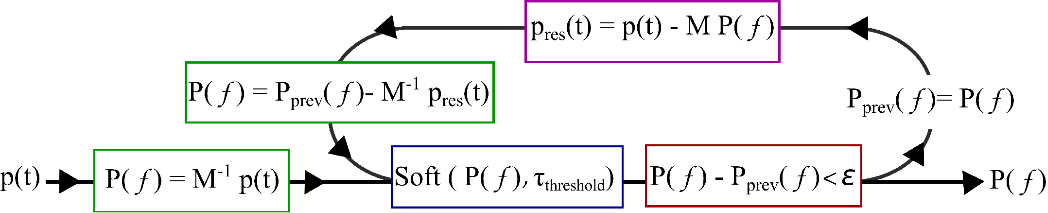}
\caption{(Color online.) Block diagram to illustrate the loop in the soft thresholding algorithm. The low-noise observerable as a function of time $p(t)$ is Fourier transformed into the frequency domain. The data is then passed through the a soft threshold, in Eq.~(\ref{eq:soft}), using $\tau_{threshold}$. The residue, $p_{res}(t)$, is calculated $p_{res}(t)=p(t) - MP(f)$. $P(f)$ is then updated by $P(f) = P_{prev}(f)-M^{-1}$res. The loop is then repeated updating $P(f)$ at each iteration until $P(f)$ converges.}
\label{fig:block}
\end{figure*}

We now briefly review compressive sensing. For a more exhaustive review, Rice University provides resources on compressive sensing~\cite{rice}. Two common techniques in compressive sensing are analogous to the two filters that reduced the counting statistics noise of the observable in the time domain as described above. In Sec III, we will use the soft threshold algorithm (that is similar to the thresholding filter described above) to remove counting statistics noise from the observable as a function of time and solve for the observable in the frequency domain. The other technique that is used in compressive sensing is the so called match pursuit~\cite{needell2009} that limits the number of frequencies, which is similar to the low-pass filter. Compressive sensing is able to extract the delta function peaks because of the sparsity of the low-noise observable in the frequency of the time domain, even if the signal as a function of time is contaminated with noise. However, the delta function peaks in the frequency domain are broadened once decoherence is added to the low-noise observable as a function of time. If the delta function peaks become too broad due to decoherence, the condition of sparsity in the frequency domain is no longer met, and the compressive sensing approach does not work as well.    

Compressive sensing solves for the s-sparse Fourier transform of the observable in the frequency domain, $P(f)$, by minimizing the following equation  
\begin{equation}
	\text{min } \left\lbrace\frac{1}{2}\left(||p(t) - M P(f)||_{l_2}\right)^2 + \tau_{threshold} ||P(f)||_{l_1}\right\rbrace.
	\label{eq:robustnoise} 
\end{equation}
where $p(t)$ is a vector of the measured observable as a function of time with decoherence and noise and $M$ is the inverse partial discrete Fourier transform matrix
\begin{equation}
	M =
	\begin{pmatrix}
		\mathcal{F}^{-1}_{1,1} & \mathcal{F}^{-1}_{1,2} & \cdots & \mathcal{F}^{-1}_{1,N_{step}} \\
		\mathcal{F}^{-1}_{2,1} & \mathcal{F}^{-1}_{2,2} & \cdots & \mathcal{F}^{-1}_{2,N_{step}}\\
		\vdots				   & \vdots					& \ddots & \vdots\\ 					
		\mathcal{F}^{-1}_{M_{step},1} & \mathcal{F}^{-1}_{M_{step},2} & \cdots & \mathcal{F}^{-1}_{M_{step},N_{step}}
	\end{pmatrix}.
\end{equation}
This matrix neglects time equal to zero as well as frequency equal to zero. We observed that by neglecting the DC frequency and the initial time, $t = 0$, reduced the probability of compressive sensing to produce spurious delta function peaks. The $l_p$ norm is defined as
\begin{equation}
	||P(f)||_{l_p} = \left(\sum_k^N |P(f_k)|^{p}\right)^{1/p}
\end{equation}
where $p = 0, 1, 2, \dots, \infty$. In Eq.~(\ref{eq:robustnoise}), the first term measures how accurate the solution $P(f)$ matches the observable as a function of time, $p(t)$, and the second measures how sparse the solution $P(f)$ is. The $\tau_{threshold}$ parameter balances between the sparsity and accuracy of $P(f)$. 

The solution $P(f)$ is found by using the sparse reconstruction by separable approximation (SpaRSa) framework~\cite{wright2009} that applies a soft threshold, Soft$(P(f), \tau_{threshold})$, 
\begin{equation}
	\text{Soft}(P(f), \tau_{threshold}) = \text{max}\left\lbrace\frac{|P(f)| - \tau_{threshold}}{|P(f)|}P(f), 0\right\rbrace.
	\label{eq:soft}
\end{equation}
The soft thresholding algorithm begins by Fourier transforming the measured observable with decoherence and noise to the frequency domain $M^{-1}p(t) = P(f)$. Then the loop starts with applying the soft thresholding to $P(f)$. The resulting $P(f)$ after the soft thresholding is transformed to the time domain and the residual is found between the measured observable as a function of time and $P(f)$, $p_{res}(t) = p(t) - MP(f)$. The residual is Fourier transformed into the frequency domain and $P(f)$ is updated. This loop is then repeated updating $P(f)$ at each iteration until the relative changes of $P(f)$ between iterations is less than $\epsilon$ as depicted schematically in Fig.~\ref{fig:block}. The threshold filter used to reduce the counting statistics noise of the observable as a function of time in Fig.~\ref{fig:compression} is analogous to running the soft threshold. The SparSa framework allows $\tau_{threshold}$ to vary at each iteration of the soft threshold and directs $\tau_{threshold}$ so that the soft thresholding algorithm efficiently converges to the $P(f)$ with the highest probability of being correct. Although the $\tau_{threshold}$ parameter can vary, an initial $\tau_{threshold}$ is needed. In computer science, the classic approach to finding an optimal guess for $\tau_{threshold}$ is called cross-validation~\cite{ward2009}. 

Cross-validation randomly assigns the low-noise observable as a function into two data sets with equal numbers of elements. One of the data sets is called the training set and the other is the test set. We apply the SparSa framework onto the training set with a trial $\tau_{threshold}$ chosen from the interval $[0.05, 1] \times ||P(f)||_{l_\infty}$, where $||P(f)||_{l_\infty} = $ max$\lbrace|P(f)|\rbrace$. The resulting $P(f)$ is Fourier transformed to the time domain and compared to the test set. The $\tau_{threshold}$ with the lowest difference between the $MP(f)$ and the test set is then used for the SparSa framework applied on the complete low-noise observable.  

\section{III. Results}
\begin{figure}
	\centering
	\includegraphics[scale=0.3]{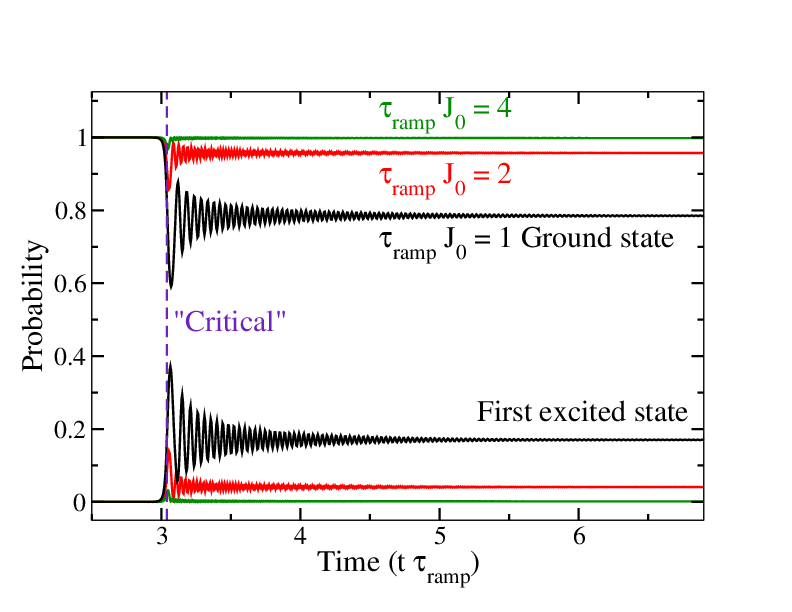}
	\caption{(Color online.) Probability to be in the instantaneous ground state (top) and first coupled excited state (bottom) as a function of time for $N_{part.} = 400$. The probability to create excitations increases as $\tau_{ramp}$ in Eq.~(\ref{eq:magnetic}) is decreased, depicted here with $\tau_{ramp} J_0= 4$ (green), $2$ (red), $ 1$ (black).}
	\label{fig:probtrans}
\end{figure}

We present a numerical example to illustrate the proposed spectroscopy protocol. We use $J_0 = 10$kHz as the energy unit. We work with a system of spins of $N_{part.} = 400$ spins. The minimum energy gap between the first coupled excited state and the ground state occurs at a ``critical" transverse magnetic field strength of $B^{(x)}(t)/J_0 = 0.4783$. For the infinite-range Ising model the probability to create excitations from the ground state increases as the transverse magnetic field ramping time rate, $\tau_{ramp}$, is decreased, as depicted in Fig.~\ref{fig:probtrans}. When $\tau_{ramp} J_0=4$ in Eq.~(\ref{eq:magnetic}) the probability to create excitations is nearly zero after the minimum energy gap. As $\tau_{ramp}$ is decreased to $2$ more excitations are created in comparison to $4$.  
We work with a $\tau_{ramp} J_0= 2$, the red line in Fig.~\ref{fig:probtrans}. The frequency accuracy is set to $\delta f = f_N/1024$ and the observable is measured at $200$ time steps. 
\begin{figure}
	\begin{center}
		\includegraphics[scale=0.3]{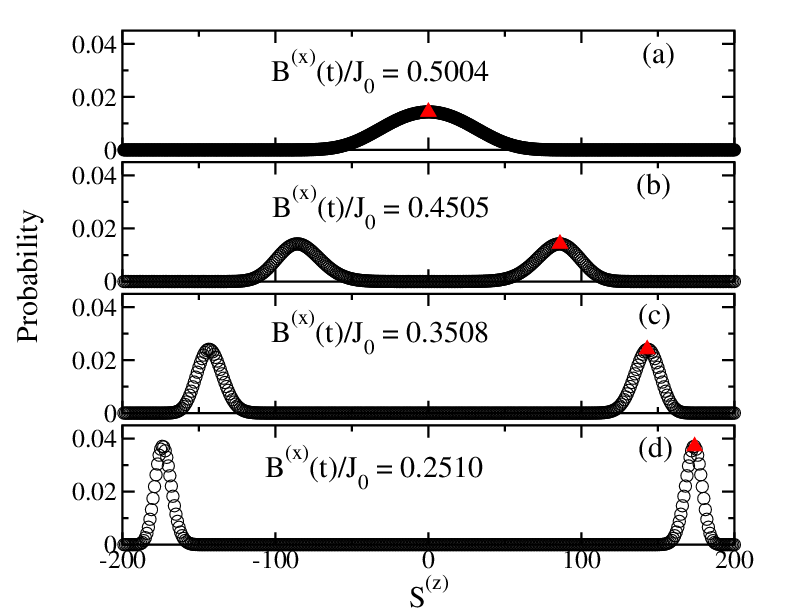}
		\caption{(Color online) Example probability of the product state shown at four different transverse magnetic field strengths (black circles). The highest probable state of the product state is signified by the red triangles for the four different transverse magnetic field strengths. The highest probable state is found to be (a) $S^{(z)}/ = 0$ for $B{(x)}/J_0 =0.5004$, (b) $S^{(z)}/ = 86$ for $B{(x)}/J_0 = 0.4505$, (c) is $S^{(z)}/ = 143$ for $B{(x)}/J_0 = 0.3508$, and (d) is $S^{(z)}/ = 178$ for $B{(x)}/J_0 = 0.2510$}
		\label{fig:initprob}
	\end{center}
\end{figure}

\begin{figure*}	
	\begin{center}
	\begin{tabular}{c c}
		\includegraphics[scale=0.3]{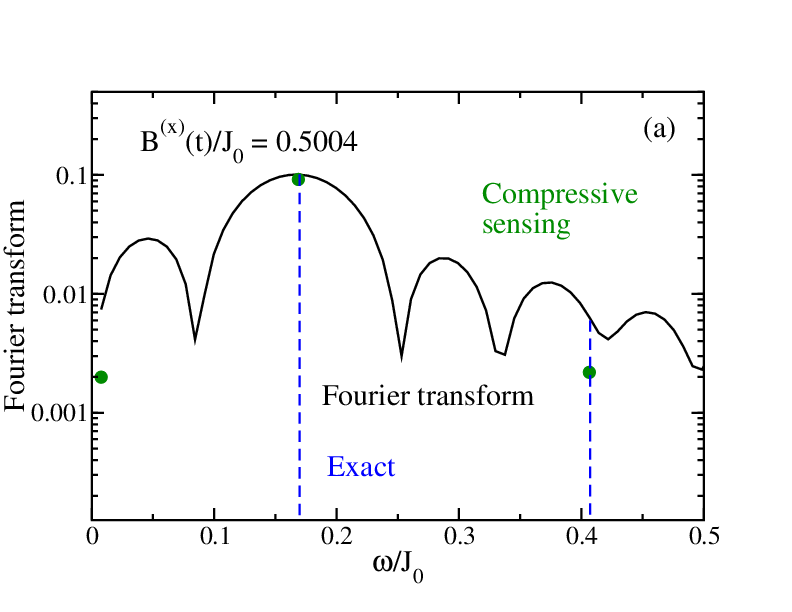} & \includegraphics[scale=0.3]{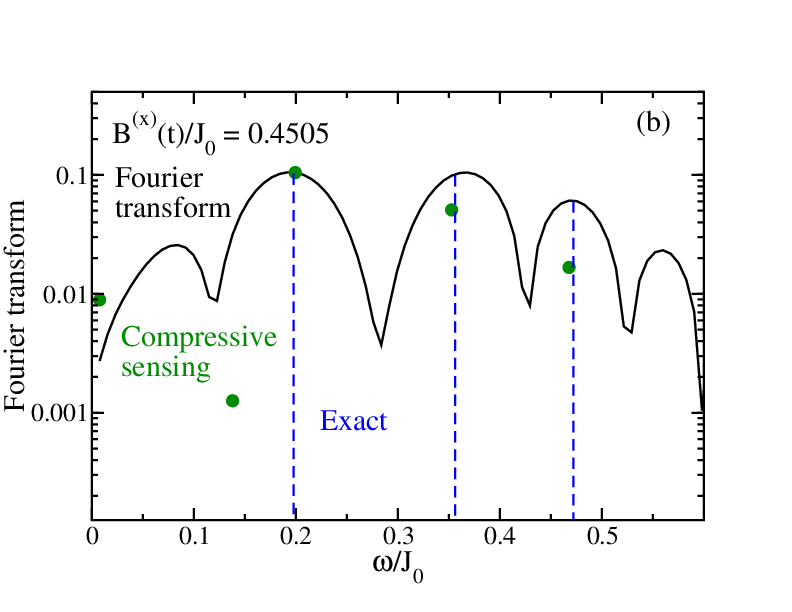}\\
		\includegraphics[scale=0.3]{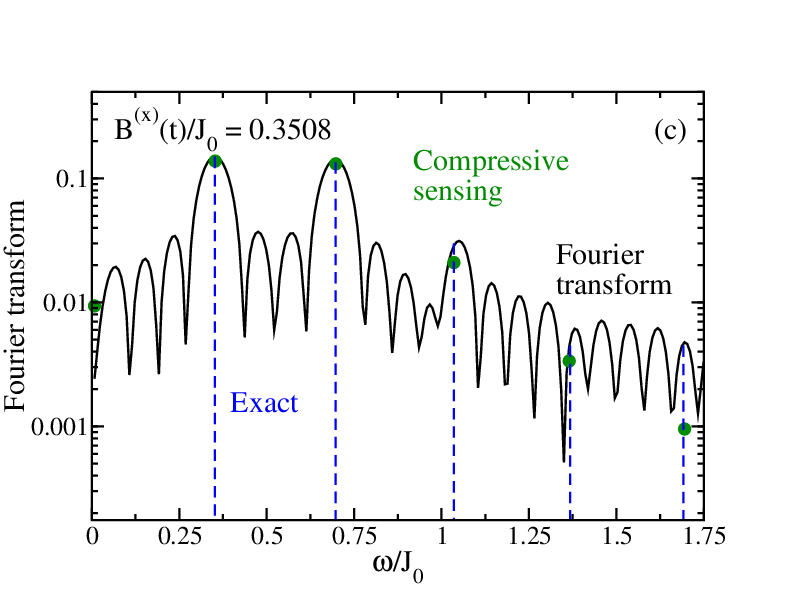} & \includegraphics[scale=0.3]{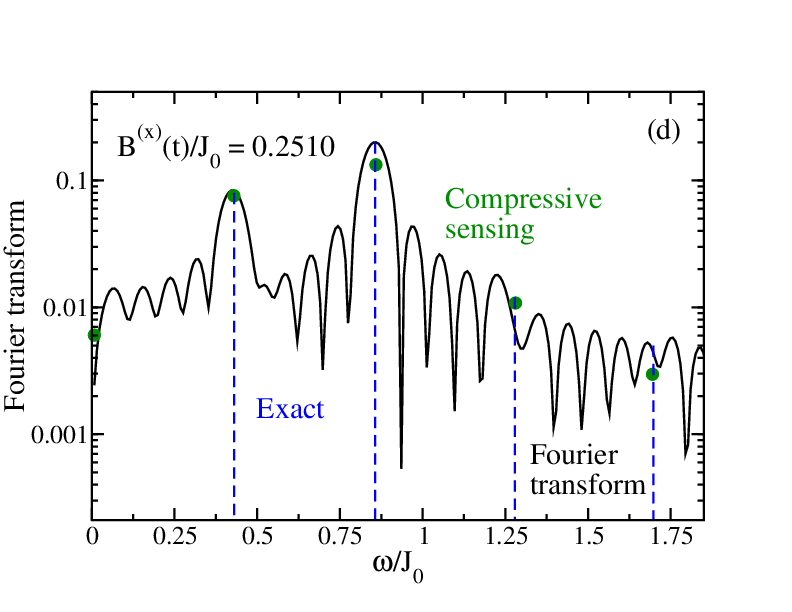}
	\end{tabular}
		\caption{(Color online.) Performing a Fourier transform (black line) on the noiseless numerical time evolution of the observable shown as a black line in Fig.~\ref{fig:signal} for the four transverse magnetic fields. The Fourier transform broadens the delta function due to the finite time interval having $200$ time steps. Using compressive sensing the delta function peaks are extracted (green filled circles). The delta function peaks that are found by the compressive sensing algorithm are compared to the adiabatic energy differences of the coupled excited states with the ground state at the four different transverse magnetic fields, where the adiabatic energies are the blue dashed line.}
		\label{fig:noiseless}
	\end{center}
	
\end{figure*} 

The observable we chose to measure while the transverse magnetic field is held constant is the highest probable state of the product state at $t_{stop}$. We also tested using observables like the average magnetization but found they typically generate more spurious peaks when processed with the compressive sensing algorithm. In Fig.~\ref{fig:initprob}, we plot four example probabilities of the product state at $B^{(x)}(t)/J_0 = 0.5004$, $0.4505$, $0.3508$ and $0.2510$ and the states $S^{(z)}/= 0$, $86$, $143$ and $174$ have the highest probability for each transverse magnetic field, respectively.

\subsection{A. Noiseless}

The noiseless numerical time evolution of the observable is plotted in Fig.~\ref{fig:signal} at $4$ different transverse magnetic field strengths, as in Fig.~\ref{fig:initprob}. Figure ~\ref{fig:signal}(a) shows the observable as a function of time before the ``critical" transverse magnetic field, $B^{(x)}/J_0 = 0.5004$. There is a small amplitude oscillation in the observable as a function of time due to a low probability to be in the coupled excited state as expected from Eq.~(\ref{eq:oscillations}). Once the transverse magnetic field has passed the ``critical" transverse magnetic field strength, as depicted in Fig.~\ref{fig:signal}(b-c), more excitations are created to the coupled excited states resulting in larger amplitude oscillations of the superposition of multiple excited states with the ground state.    

The observable in Fig.~\ref{fig:signal} is Fourier transformed to the frequency domain to determine frequencies of the oscillations. The frequencies of the oscillations are the energy differences of the coupled excited states, as found in Eq.~(\ref{eq:oscillations}). Due to the finite time interval, the Fourier transform of the delta function peaks of the observable have been broadened in the frequency domain, as seen in Fig.~\ref{fig:noiseless}. Using compressive sensing the delta function peaks are recovered. Fig.~\ref{fig:noiseless}(a) has delta function peaks at frequencies of the energy differences of the first and second coupled excited states with the ground state. When the transverse magnetic field is stopped immediately after the ``critical" transverse magnetic field strength, as shown in Fig.~\ref{fig:noiseless}(b), the energy difference of the lowest three coupled excited states with the ground state are found. However, a spurious peak appears at a frequency lower than the first lowest-lying coupled excited state. Further decreasing the transverse magnetic field, the energy differences of the lowest four coupled excited states with the ground state are extracted, as found in Fig.~\ref{fig:noiseless}(c) and (d).

\begin{figure}
\centering
	\begin{tabular}{c}
		\includegraphics[scale=0.175]{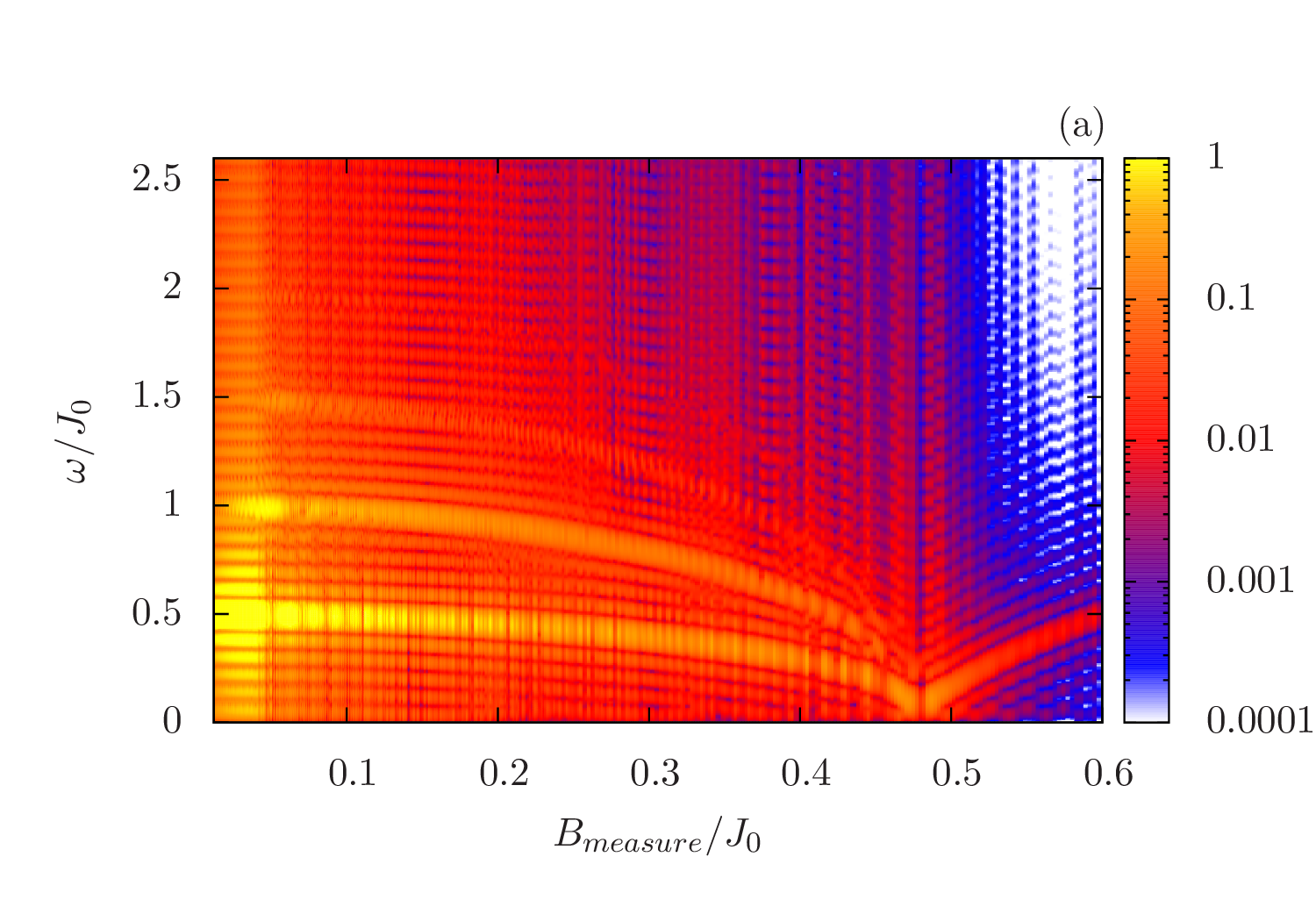}\\
		\includegraphics[scale=0.3]{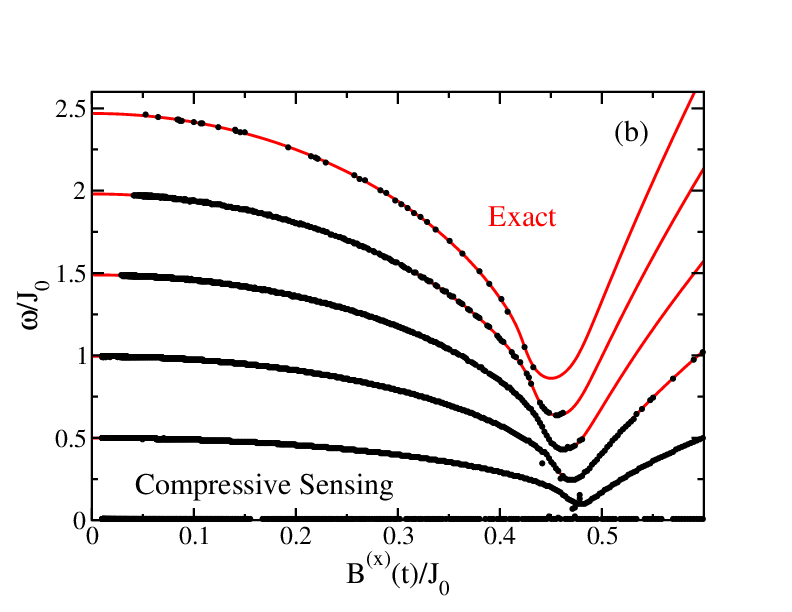}\\
	\end{tabular}
\caption{(Color online.) Energy spectra extracted by either applying the (a) partial discrete Fourier transform, $M^{-1}$, or compressive sensing (b) on the noiseless numerical time evolution observable as a function of the transverse magnetic field. In panel a, due to applying the noiseless numerical time evolution by the partial discrete Fourier transform the delta function peaks of the lowest four lying coupled excited states have become broadened and ``ringing" is obscuring the third and fourth lowest-lying coupled excited states. Alternatively by applying compressive sensing, the peaks (black dots) recovering the delta function peaks, as depicted in panel b. The four lowest-lying coupled excited states essentially lie on top of the adiabatic energy difference (red lines) when the transverse magnetic field is less than the ``critical" transverse magnetic field. The fifth lowest lying coupled excited state is extracted at a few of the transverse magnetic field strengths. Spurious delta function peaks occur at transverse magnetic fields near the ``critical" transverse magnetic field.}
\label{fig:nlspectra}
\end{figure}

We can produce energy spectra as a function of the transverse magnetic field, as shown in Fig.~\ref{fig:nlspectra}, by applying the partial discrete Fourier transform or compressive sensing to the time evolution of the observable at different transverse magnetic field strengths. Excitations to the first coupled excited occur in the entire interval of the transverse magnetic field plotted in Fig.~\ref{fig:nlspectra} using both the partial discrete Fourier transform and compressive sensing. However at higher transverse magnetic field strengths, the excitations to the second coupled excited state also occur before the ``critical" transverse magnetic field is observed when using compressive sensing, as depicted in Fig.~\ref{fig:nlspectra}(b). After the ``critical" transverse magnetic field, excitations to the higher coupled excited states are found when applying the partial discrete Fourier transform or compressive sensing. In Fig.~\ref{fig:nlspectra}(a), the lowest three lying coupled excited states are found (while the fourth can be faintly seen). However, the peaks of the lowest coupled excited states are broadened and the third and fourth lowest-lying coupled excited states are nearly indistinguishable to the background noise, or ``ringing", which is an artifact of applying the partial discrete Fourier transform to signal that does not have complete oscillations. By applying compressive sensing to the observable the lowest four lying coupled excited states are found (at a few transverse magnetic field strengths the fifth coupled excited state is also observed). 

\subsection{B. Simulated data}

\begin{figure}
	\begin{center}
		\includegraphics[scale=0.3]{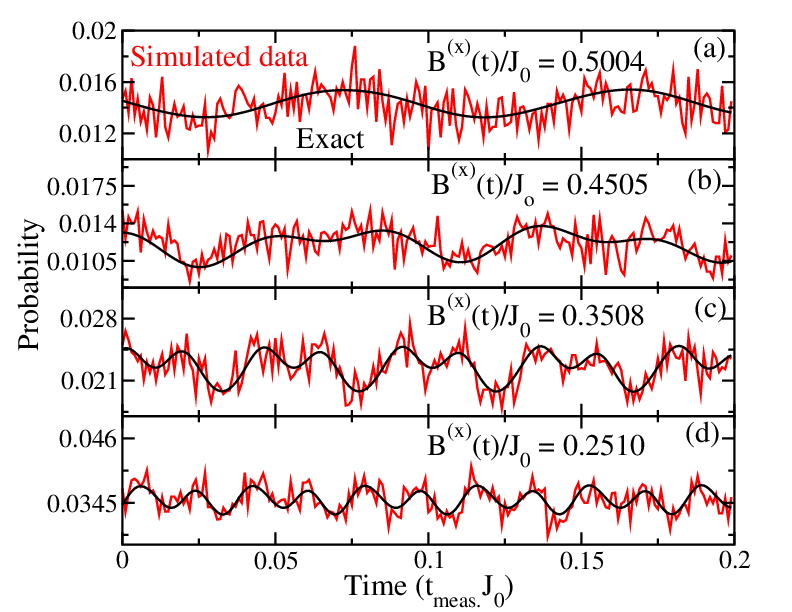}
		\caption{(Color online.) Time evolution of the observable as a function of time at four different transverse magnetic fields where (a)$B^{(x)}(t)/J_0 = 0.5004$, (b) $B^{(x)}(t)/J_0 = 0.4505$, (c) $B^{(x)}(t)/J_0 = 0.3508$, and (d) $B^{(x)}(t)/J_0 = 0.2510$. The numerical time evolution of the observable is in black and the simulated data, where the error is added to the observable as a function of time due to decoherence and counting statistics, is in red. The number of time steps is shown is $200$.}
		\label{fig:signal}
	\end{center}
\end{figure} 

\begin{figure*}	
	\begin{center}
	\begin{tabular}{c c}
		\includegraphics[scale=0.3]{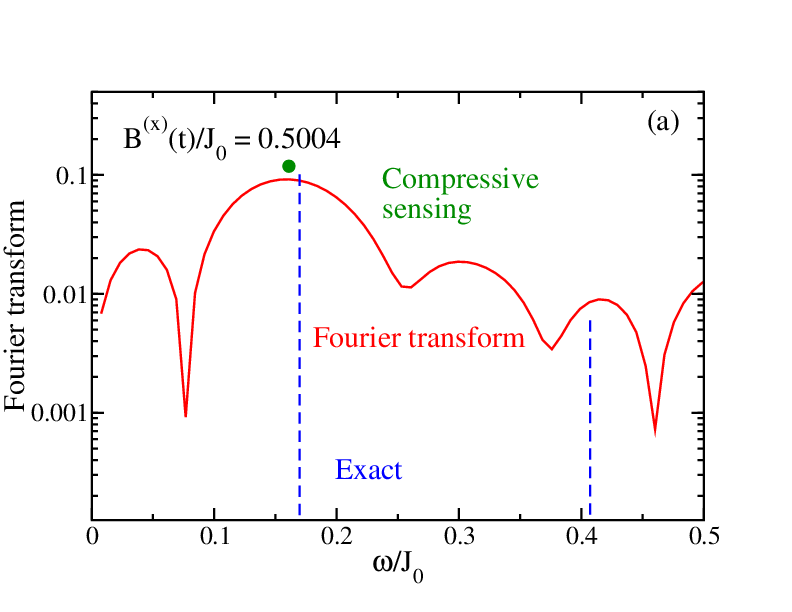} & \includegraphics[scale=0.3]{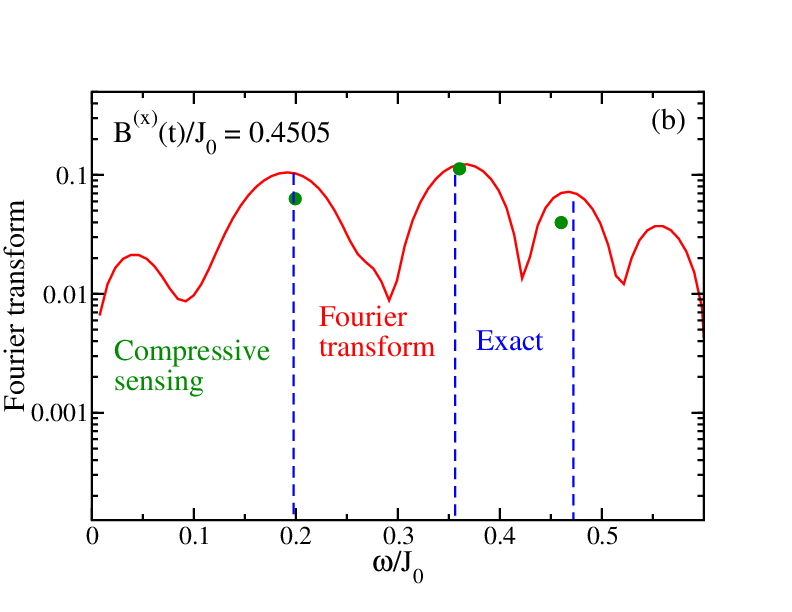}\\
		\includegraphics[scale=0.3]{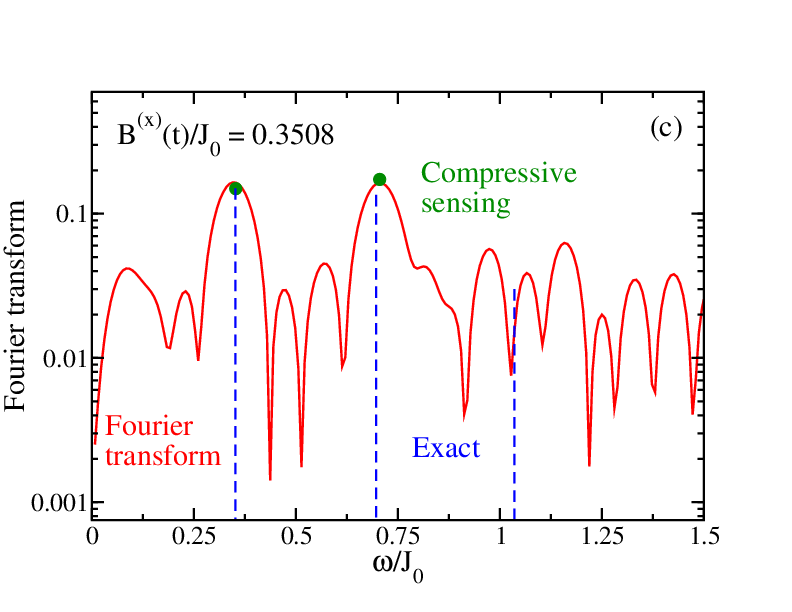} & \includegraphics[scale=0.3]{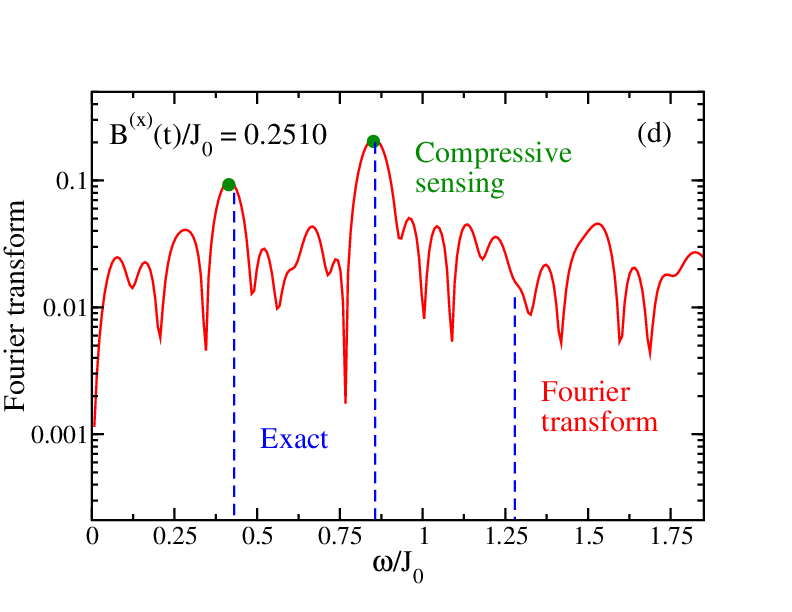}
	\end{tabular}
		\caption{(Color online.) Fourier transforming the simulated data (red) there are discernable peaks that could be extracted and compared to the adiabatic energy differences. By applying compressive sensing to the simulated data (solid green circles) the delta function peaks are extracted from the Fourier transform of the simulated data. The adiabatic energy differences, $E_n -E_0$, are plotted as blue dashed lines to compare to the peaks observed in the Fourier transform and the delta functions extracted by the compressive sensing algorithm when applied to the simulated data.}
		\label{fig:nFT}
	\end{center}
\end{figure*} 

The simulated experimental data are produced by adding decoherence, which is modeled by Eq.~(\ref{eq:decoherence}), and counting statistics error, as defined in Eq.~(\ref{eq:poisson}), to the observable as a function of time, as depicted by red lines in Fig.~\ref{fig:signal}. We used $N_{meas.} = 10,000$ at each time step to add the Poisson noise. 

In Fig.~\ref{fig:nFT}, we show four examples to compare the Fourier transform to the compressive sensing applied to the simulated data from Fig.~\ref{fig:signal}. The effects of adding the decoherence and counting statistics error to the noiseless data creates a noise floor that nearly obscures the delta function peaks associated with energy differences, as depicted in Fig.~\ref{fig:nFT}, when the Fourier transform is applied. The Fourier transform of the simulated data could be refined by extracting the peaks. However, doing this processing will result in multiple spurious peaks. The compressive sensing is able to extract energy differences of the lowest two coupled excited states [in Fig.~\ref{fig:nFT}(b) the third coupled excited state was also found] that are near the expected adiabatic energy differences. 

\begin{table}
\centering
\begin{tabular}{ |l|l|l| }
\hline
Transverse magnetic field & Adiabatic $(J_0)$ & Ave. $\pm$ STD $(J_0)$\\ \hline 
$B^{(x)}/J_0 = 0.5004$  & 0.1698 & 0.16935 $\pm$ 0.0012\\ \hline
\multirow{3}{*}{$B^{(x)}/J_0 = 0.4482$} & 0.2039 &  0.2073 $\pm$ 0.0023  \\
 & 0.3740 &  0.3708  $\pm$ 0.0019 \\
 & 0.4920 &  0.4922  $\pm$ 0.0044 \\ \hline
\multirow{2}{*}{$B^{(x)}/J_0 = 0.3976$} & 0.2961 &  0.2971 $\pm$ 0.00098 \\
 & 0.5822 & 0.5809 $\pm$ 0.0016\\ \hline
\multirow{2}{*}{$B^{(x)}/J_0 =  0.3561$} & 0.3463 & 0.3473 $\pm$ 0.0012 \\
 & 0.6858 & 0.6884 $\pm$ 0.0026\\ \hline
\multirow{2}{*}{$B^{(x)}/J_0 = 0.3065$} & 0.391797 & 0.3935  $\pm$ 0.0012 \\
 & 0.778706 & 0.7771 $\pm$ 0.0026\\ \hline
\multirow{2}{*}{$B^{(x)}/J_0 = 0.2560$} & 0.427073 & 0.4278  $\pm$ 0.00053 \\
 & 0.850257 & 0.8495 $\pm$ 0.0020\\ \hline
\multirow{2}{*}{$B^{(x)}/J_0 = 0.2055$} & 0.453929 & 0.4527 $\pm$ 0.00095 \\
 & 0.904569 & 0.90495 $\pm$ 0.0032\\ \hline
\multirow{2}{*}{$B^{(x)}/J_0 = 0.1545$} & 0.473939 & 0.4742 $\pm$ 0.00055 \\
 & 0.944969 & 0.9438 $\pm$ 0.0018 \\ \hline
\multirow{2}{*}{$B^{(x)}/J_0 = 0.1046$} & 0.487537 & 0.4880 $\pm$ 0.00060 \\
 & 0.972398 & 0.9723 $\pm$ 0.0020\\ \hline
\multirow{3}{*}{$B^{(x)}/J_0 = 0.0625$} & 0.494777 & 0.4948 $\pm$ 0.00066\\
& 0.986993 & 0.9894 $\pm$ 0.0033\\
 & 1.47665 & 1.4743 $\pm$ 0.0077\\ \hline
\end{tabular}
\caption{Energy differences of the infinite-range transverse field Ising model compared to the  average and standard deviation of the extracted observable in the frequency domain by applying the compressive sensing algorithm to $100$ different cases of the simulated data. For each of the $100$ simulated data cases the counting statistics noise is different but the decoherence error is the same, at $10$ transverse magnetic field strengths.}
\label{tab:stats}
\end{table}
 
The effects of the counting statistics and decoherence errors are studied more quantitatively by calculating the average and standard deviation of the delta function peaks produced from processing the simulated data with the compressive sensing algorithm at $10$ different transverse magnetic field strengths. The average and standard deviation are calculated at each transverse magnetic field by using the compressive sensing on $100$ cases with different counting statistics noise applied to each case and having the same decoherence time, $\tau_{d} J_0= 25$. In the majority of the transverse magnetic field strengths in Table~\ref{tab:stats} the compressive sensing was able to extract frequencies that are within $2$ digits of accuracy and standard deviations are on the order of $0.001$'s.  

There are spurious delta function peaks that appear in the $100$ different cases of the simulated data at the different transverse magnetic fields. These spurious delta function peaks usually have high frequencies in respect with the frequencies that correspond to energy differences. The occurrence of spurious delta function peaks at high frequencies is low so they are neglected in the statistical analysis. There are spurious delta functions with frequencies lower than the first lowest-lying coupled excited states included in the statistical analysis due to their consistent occurrence at three different transverse magnetic fields. 

\begin{figure}
\centering
\includegraphics[scale=0.3]{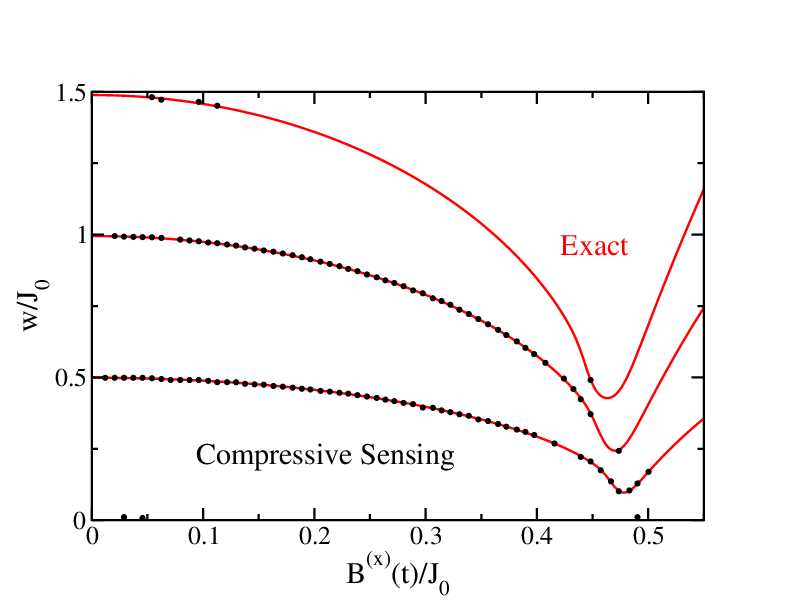}
\caption{(Color online.) Average delta function peaks from applying compressive sensing to $100$ different cases of the simulated data as a function of the transverse magnetic field strength (solid black circles). The two lowest-lying coupled excited states are identified when comparing to the adiabatic energy differences (red dashed lines). The average delta function peaks have small deviations from the adiabatic energy differences as shown in in ~\ref{tab:stats}. There low frequency spurious delta function peaks that occur at three transverse magnetic fields.}
\label{fig:averageCS}
\end{figure}

The average extracted delta function peaks found by compressive sensing are plotted as a function of the transverse magnetic field in Fig.~\ref{fig:averageCS}. At low transverse magnetic field the lowest two coupled excited states can be identified. 

\section{IV. Conclusion}
In this work we have proposed a spectroscopy protocol that diabatically ramps the transverse magnetic field to create excitations. By diabatically ramping and then holding the transverse magnetic field, the energy spectra can be extracted by measuring a low-noise observable as a function of time and then signal processing the data. We explored our protocol by simulating data for the infinite-range transverse field Ising model. By using compressive sensing, the number of time steps needed for the signal processing is sharply reduced and the spectroscopy protocol becomes experimentally feasible with current experimental setups. This occurs because compressive sensing is robust against counting statistics errors. However, compressive sensing is not robust against errors due to decoherence, which can result in spurious delta functions peaks. We find by using compressive sensing on the noiseless numerical time evolution of the observable that a number of lowest lying energy states can be extracted. When counting statistics and decoherence errors are added to the observable as a function of time, compressive sensing can extract fewer low lying energy states as the transverse magnetic field approaches zero. At high transverse magnetic field, the probability to create excitations is too low with respect to the errors added and this results in spurious delta function peaks that are not associated with any energy levels. We hope our protocol will be used in current simulations to extract interesting many-body spectra.  
 
\section{Acknowledgments}
We thank Crystal Senko and Chris Monroe for valuable discussions. J. K. F. and B. Y. acknowledge support from the National Science Foundation under grant number PHY-1314295. J. K. F. also acknowledges support from the McDevitt bequest at Georgetown University. B. Y. acknowledges support from the Achievement Rewards for College Students Foundation. W. C. C. acknowledges support from the U.S. Air Force Ofﬁce of Scientiﬁc Research Young Investigator Program under award number FA9550-13-1-0167 

\bibliographystyle{apsrev}
\bibliography{UniformInteraction}

\begin{thebibliography}{35}
\expandafter\ifx\csname natexlab\endcsname\relax\def\natexlab#1{#1}\fi
\expandafter\ifx\csname bibnamefont\endcsname\relax
  \def\bibnamefont#1{#1}\fi
\expandafter\ifx\csname bibfnamefont\endcsname\relax
  \def\bibfnamefont#1{#1}\fi
\expandafter\ifx\csname citenamefont\endcsname\relax
  \def\citenamefont#1{#1}\fi
\expandafter\ifx\csname url\endcsname\relax
  \def\url#1{\texttt{#1}}\fi
\expandafter\ifx\csname urlprefix\endcsname\relax\def\urlprefix{URL }\fi
\providecommand{\bibinfo}[2]{#2}
\providecommand{\eprint}[2][]{\url{#2}}

\bibitem[{\citenamefont{Sachdev}(1999)}]{sachdev1999}
\bibinfo{author}{\bibfnamefont{S.}~\bibnamefont{Sachdev}},
  \emph{\bibinfo{title}{Quantum Phase Transitions}}
  (\bibinfo{publisher}{Cambridge University Press}, \bibinfo{year}{1999}).

\bibitem[{\citenamefont{Diep}(2005)}]{diep2005}
\bibinfo{author}{\bibfnamefont{H.~T.} \bibnamefont{Diep}},
  \emph{\bibinfo{title}{Frustrated Spin Systems}} (\bibinfo{publisher}{World
  Scientific}, \bibinfo{year}{2005}).

\bibitem[{\citenamefont{Moessner and Ramirez}(2006)}]{moessner2006}
\bibinfo{author}{\bibfnamefont{R.}~\bibnamefont{Moessner}} \bibnamefont{and}
  \bibinfo{author}{\bibfnamefont{A.~P.} \bibnamefont{Ramirez}},
  \bibinfo{journal}{Physics Today} \textbf{\bibinfo{volume}{59}},
  \bibinfo{pages}{24} (\bibinfo{year}{2006}).

\bibitem[{\citenamefont{Feynman}(1981)}]{feynman1981}
\bibinfo{author}{\bibfnamefont{R.~P.} \bibnamefont{Feynman}},
  \bibinfo{journal}{Int. J. Theory. Phys.} \textbf{\bibinfo{volume}{21}},
  \bibinfo{pages}{467} (\bibinfo{year}{1981}).

\bibitem[{\citenamefont{Friedenauer et~al.}(2008)\citenamefont{Friedenauer,
  Schmitz, Glueckert, Porras, and Schaetz}}]{friedenauer2008}
\bibinfo{author}{\bibfnamefont{A.}~\bibnamefont{Friedenauer}},
  \bibinfo{author}{\bibfnamefont{H.}~\bibnamefont{Schmitz}},
  \bibinfo{author}{\bibfnamefont{J.~T.} \bibnamefont{Glueckert}},
  \bibinfo{author}{\bibfnamefont{D.}~\bibnamefont{Porras}}, \bibnamefont{and}
  \bibinfo{author}{\bibfnamefont{T.}~\bibnamefont{Schaetz}},
  \bibinfo{journal}{Nat. Phys.} \textbf{\bibinfo{volume}{4}},
  \bibinfo{pages}{757} (\bibinfo{year}{2008}).

\bibitem[{\citenamefont{Kim et~al.}(2009)\citenamefont{Kim, Chang, Islam,
  Korenblit, Duan, and Monroe}}]{kim2009}
\bibinfo{author}{\bibfnamefont{K.}~\bibnamefont{Kim}},
  \bibinfo{author}{\bibfnamefont{M.-S.} \bibnamefont{Chang}},
  \bibinfo{author}{\bibfnamefont{R.}~\bibnamefont{Islam}},
  \bibinfo{author}{\bibfnamefont{S.}~\bibnamefont{Korenblit}},
  \bibinfo{author}{\bibfnamefont{L.-M.} \bibnamefont{Duan}}, \bibnamefont{and}
  \bibinfo{author}{\bibfnamefont{C.}~\bibnamefont{Monroe}},
  \bibinfo{journal}{Phys. Rev. Lett.} \textbf{\bibinfo{volume}{103}},
  \bibinfo{pages}{120502} (\bibinfo{year}{2009}).

\bibitem[{\citenamefont{Kim et~al.}(2010)\citenamefont{Kim, Chang, Korenblit,
  Islam, Edwards, Freericks, Lin, Duan, and Monroe}}]{kim2010}
\bibinfo{author}{\bibfnamefont{K.}~\bibnamefont{Kim}},
  \bibinfo{author}{\bibfnamefont{M.-S.} \bibnamefont{Chang}},
  \bibinfo{author}{\bibfnamefont{S.}~\bibnamefont{Korenblit}},
  \bibinfo{author}{\bibfnamefont{R.}~\bibnamefont{Islam}},
  \bibinfo{author}{\bibfnamefont{E.~E.} \bibnamefont{Edwards}},
  \bibinfo{author}{\bibfnamefont{J.~K.} \bibnamefont{Freericks}},
  \bibinfo{author}{\bibfnamefont{G.-D.} \bibnamefont{Lin}},
  \bibinfo{author}{\bibfnamefont{L.-M.} \bibnamefont{Duan}}, \bibnamefont{and}
  \bibinfo{author}{\bibfnamefont{C.}~\bibnamefont{Monroe}},
  \bibinfo{journal}{Nature} \textbf{\bibinfo{volume}{465}},
  \bibinfo{pages}{590} (\bibinfo{year}{2010}).

\bibitem[{\citenamefont{Islam et~al.}(2011)\citenamefont{Islam, Edwards, Kim,
  Korenblit, Noh, Carmichael, Lin, Duan, Wang, Freericks et~al.}}]{islam2011}
\bibinfo{author}{\bibfnamefont{R.}~\bibnamefont{Islam}},
  \bibinfo{author}{\bibfnamefont{E.~E.} \bibnamefont{Edwards}},
  \bibinfo{author}{\bibfnamefont{K.}~\bibnamefont{Kim}},
  \bibinfo{author}{\bibfnamefont{S.}~\bibnamefont{Korenblit}},
  \bibinfo{author}{\bibfnamefont{C.}~\bibnamefont{Noh}},
  \bibinfo{author}{\bibfnamefont{H.}~\bibnamefont{Carmichael}},
  \bibinfo{author}{\bibfnamefont{G.-D.} \bibnamefont{Lin}},
  \bibinfo{author}{\bibfnamefont{L.-M.} \bibnamefont{Duan}},
  \bibinfo{author}{\bibfnamefont{C.-C.~J.} \bibnamefont{Wang}},
  \bibinfo{author}{\bibfnamefont{J.~K.} \bibnamefont{Freericks}},
  \bibnamefont{et~al.}, \bibinfo{journal}{Nature Comm.}
  \textbf{\bibinfo{volume}{2}}, \bibinfo{pages}{1374} (\bibinfo{year}{2011}).

\bibitem[{\citenamefont{Britton et~al.}(2012)\citenamefont{Britton, Sawyer,
  Keith, Wang, Freericks, Biercuk, Uys, and Bollinger}}]{britton2012}
\bibinfo{author}{\bibfnamefont{J.~W.} \bibnamefont{Britton}},
  \bibinfo{author}{\bibfnamefont{B.~C.} \bibnamefont{Sawyer}},
  \bibinfo{author}{\bibfnamefont{A.~C.} \bibnamefont{Keith}},
  \bibinfo{author}{\bibfnamefont{C.-C.~J.} \bibnamefont{Wang}},
  \bibinfo{author}{\bibfnamefont{J.~K.} \bibnamefont{Freericks}},
  \bibinfo{author}{\bibfnamefont{M.~J.} \bibnamefont{Biercuk}},
  \bibinfo{author}{\bibfnamefont{H.}~\bibnamefont{Uys}}, \bibnamefont{and}
  \bibinfo{author}{\bibfnamefont{J.~J.} \bibnamefont{Bollinger}},
  \bibinfo{journal}{Nature} \textbf{\bibinfo{volume}{484}},
  \bibinfo{pages}{489} (\bibinfo{year}{2012}).

\bibitem[{\citenamefont{Islam et~al.}(2013)\citenamefont{Islam, Senko,
  Campbell, Korenblit, Smith, Lee, Edwards, Wang, Freericks, and
  Monroe}}]{islam2013}
\bibinfo{author}{\bibfnamefont{R.}~\bibnamefont{Islam}},
  \bibinfo{author}{\bibfnamefont{C.}~\bibnamefont{Senko}},
  \bibinfo{author}{\bibfnamefont{W.~C.} \bibnamefont{Campbell}},
  \bibinfo{author}{\bibfnamefont{S.}~\bibnamefont{Korenblit}},
  \bibinfo{author}{\bibfnamefont{J.}~\bibnamefont{Smith}},
  \bibinfo{author}{\bibfnamefont{A.}~\bibnamefont{Lee}},
  \bibinfo{author}{\bibfnamefont{E.~E.} \bibnamefont{Edwards}},
  \bibinfo{author}{\bibfnamefont{C.-C.~J.} \bibnamefont{Wang}},
  \bibinfo{author}{\bibfnamefont{J.~K.} \bibnamefont{Freericks}},
  \bibnamefont{and} \bibinfo{author}{\bibfnamefont{C.}~\bibnamefont{Monroe}},
  \bibinfo{journal}{Science} \textbf{\bibinfo{volume}{340}},
  \bibinfo{pages}{583} (\bibinfo{year}{2013}).

\bibitem[{\citenamefont{Senko et~al.}(2014)\citenamefont{Senko, Smith,
  Richerme, Lee, Campbell, and Monroe}}]{senko2013}
\bibinfo{author}{\bibfnamefont{C.}~\bibnamefont{Senko}},
  \bibinfo{author}{\bibfnamefont{J.}~\bibnamefont{Smith}},
  \bibinfo{author}{\bibfnamefont{P.}~\bibnamefont{Richerme}},
  \bibinfo{author}{\bibfnamefont{A.}~\bibnamefont{Lee}},
  \bibinfo{author}{\bibfnamefont{W.~C.} \bibnamefont{Campbell}},
  \bibnamefont{and} \bibinfo{author}{\bibfnamefont{C.}~\bibnamefont{Monroe}},
  \bibinfo{howpublished}{e-print} (\bibinfo{year}{2014}),
  \urlprefix\url{http://arxiv.org/abs/1401.5751}.

\bibitem[{\citenamefont{Lloyd}(1996)}]{lloyd1996}
\bibinfo{author}{\bibfnamefont{S.}~\bibnamefont{Lloyd}},
  \bibinfo{journal}{Science} \textbf{\bibinfo{volume}{273}},
  \bibinfo{pages}{1073} (\bibinfo{year}{1996}).

\bibitem[{\citenamefont{Farhi et~al.}(2000)\citenamefont{Farhi, Goldstone,
  Gutmann, and Sipser}}]{farhi2000}
\bibinfo{author}{\bibfnamefont{E.}~\bibnamefont{Farhi}},
  \bibinfo{author}{\bibfnamefont{J.}~\bibnamefont{Goldstone}},
  \bibinfo{author}{\bibfnamefont{S.}~\bibnamefont{Gutmann}}, \bibnamefont{and}
  \bibinfo{author}{\bibfnamefont{M.}~\bibnamefont{Sipser}},
  \bibinfo{howpublished}{e-print} (\bibinfo{year}{2000}),
  \urlprefix\url{http://arxiv.org/pdf/quant-ph/0001106v1.pdf}.

\bibitem[{\citenamefont{Richerme et~al.}(2013)\citenamefont{Richerme, Senko,
  Smith, Lee, Korenblit, and Monroe}}]{richerme2013}
\bibinfo{author}{\bibfnamefont{P.}~\bibnamefont{Richerme}},
  \bibinfo{author}{\bibfnamefont{C.}~\bibnamefont{Senko}},
  \bibinfo{author}{\bibfnamefont{J.}~\bibnamefont{Smith}},
  \bibinfo{author}{\bibfnamefont{A.}~\bibnamefont{Lee}},
  \bibinfo{author}{\bibfnamefont{S.}~\bibnamefont{Korenblit}},
  \bibnamefont{and} \bibinfo{author}{\bibfnamefont{C.}~\bibnamefont{Monroe}},
  \bibinfo{journal}{Phys. Rev. A} \textbf{\bibinfo{volume}{88}},
  \bibinfo{pages}{012334} (\bibinfo{year}{2013}).

\bibitem[{\citenamefont{Lipkin et~al.}(1964{\natexlab{a}})\citenamefont{Lipkin,
  Meshkov, and Glick}}]{lipkin1964}
\bibinfo{author}{\bibfnamefont{H.~J.} \bibnamefont{Lipkin}},
  \bibinfo{author}{\bibfnamefont{N.}~\bibnamefont{Meshkov}}, \bibnamefont{and}
  \bibinfo{author}{\bibfnamefont{A.~J.} \bibnamefont{Glick}},
  \bibinfo{journal}{Nucl. Phys.} \textbf{\bibinfo{volume}{62}},
  \bibinfo{pages}{188} (\bibinfo{year}{1964}{\natexlab{a}}).

\bibitem[{\citenamefont{Lipkin et~al.}(1964{\natexlab{b}})\citenamefont{Lipkin,
  Meshkov, and Glick}}]{meshkov1964}
\bibinfo{author}{\bibfnamefont{H.~J.} \bibnamefont{Lipkin}},
  \bibinfo{author}{\bibfnamefont{N.}~\bibnamefont{Meshkov}}, \bibnamefont{and}
  \bibinfo{author}{\bibfnamefont{A.~J.} \bibnamefont{Glick}},
  \bibinfo{journal}{Nucl. Phys.} \textbf{\bibinfo{volume}{62}},
  \bibinfo{pages}{199} (\bibinfo{year}{1964}{\natexlab{b}}).

\bibitem[{\citenamefont{Lipkin et~al.}(1964{\natexlab{c}})\citenamefont{Lipkin,
  Meshkov, and Glick}}]{glick1964}
\bibinfo{author}{\bibfnamefont{H.~J.} \bibnamefont{Lipkin}},
  \bibinfo{author}{\bibfnamefont{N.}~\bibnamefont{Meshkov}}, \bibnamefont{and}
  \bibinfo{author}{\bibfnamefont{A.~J.} \bibnamefont{Glick}},
  \bibinfo{journal}{Nucl. Phys.} \textbf{\bibinfo{volume}{62}},
  \bibinfo{pages}{211} (\bibinfo{year}{1964}{\natexlab{c}}).

\bibitem[{\citenamefont{Newman and Shulman}(1977)}]{newman1977}
\bibinfo{author}{\bibfnamefont{C.~M.} \bibnamefont{Newman}} \bibnamefont{and}
  \bibinfo{author}{\bibfnamefont{L.~S.} \bibnamefont{Shulman}},
  \bibinfo{journal}{J. Math. Phys.} \textbf{\bibinfo{volume}{18}},
  \bibinfo{pages}{23} (\bibinfo{year}{1977}).

\bibitem[{\citenamefont{Gilmore and Feng}(1978)}]{gilmore1978}
\bibinfo{author}{\bibfnamefont{R.}~\bibnamefont{Gilmore}} \bibnamefont{and}
  \bibinfo{author}{\bibfnamefont{D.~H.} \bibnamefont{Feng}},
  \bibinfo{journal}{Phys. Lett.} \textbf{\bibinfo{volume}{76B}},
  \bibinfo{pages}{26} (\bibinfo{year}{1978}).

\bibitem[{\citenamefont{Botet and Jullien}(1983)}]{botet1983}
\bibinfo{author}{\bibfnamefont{R.}~\bibnamefont{Botet}} \bibnamefont{and}
  \bibinfo{author}{\bibfnamefont{R.}~\bibnamefont{Jullien}},
  \bibinfo{journal}{Phys. Rev. B} \textbf{\bibinfo{volume}{28}},
  \bibinfo{pages}{3955} (\bibinfo{year}{1983}).

\bibitem[{\citenamefont{Trotter}(1959)}]{trotter1959}
\bibinfo{author}{\bibfnamefont{H.}~\bibnamefont{Trotter}},
  \bibinfo{journal}{Proc. Am. Math. Soc} \textbf{\bibinfo{volume}{10}},
  \bibinfo{pages}{545} (\bibinfo{year}{1959}).

\bibitem[{\citenamefont{Campbell}(1897)}]{campbell1897}
\bibinfo{author}{\bibfnamefont{J.}~\bibnamefont{Campbell}},
  \bibinfo{journal}{Proc. Lond. Math Soc.} \textbf{\bibinfo{volume}{28}},
  \bibinfo{pages}{381} (\bibinfo{year}{1897}).

\bibitem[{\citenamefont{Baker}(1902)}]{baker1902}
\bibinfo{author}{\bibfnamefont{H.}~\bibnamefont{Baker}},
  \bibinfo{journal}{Proc. Lond. Math Soc} \textbf{\bibinfo{volume}{34}},
  \bibinfo{pages}{347} (\bibinfo{year}{1902}).

\bibitem[{\citenamefont{F.Hausdorff}(1906)}]{hausdorff1906}
\bibinfo{author}{\bibnamefont{F.Hausdorff}}, \bibinfo{journal}{BerVerh Saechs
  Akad Wiss Leipzig} \textbf{\bibinfo{volume}{58}}, \bibinfo{pages}{19}
  (\bibinfo{year}{1906}).

\bibitem[{\citenamefont{Alverman and Fehske}(2011)}]{alverman2011}
\bibinfo{author}{\bibfnamefont{A.}~\bibnamefont{Alverman}} \bibnamefont{and}
  \bibinfo{author}{\bibfnamefont{H.}~\bibnamefont{Fehske}},
  \bibinfo{journal}{J. Comput. Phys.} \textbf{\bibinfo{volume}{230}},
  \bibinfo{pages}{5930} (\bibinfo{year}{2011}).

\bibitem[{\citenamefont{Alverman et~al.}(2012)\citenamefont{Alverman, Fehske,
  and Littlewood}}]{alverman2012}
\bibinfo{author}{\bibfnamefont{A.}~\bibnamefont{Alverman}},
  \bibinfo{author}{\bibfnamefont{H.}~\bibnamefont{Fehske}}, \bibnamefont{and}
  \bibinfo{author}{\bibfnamefont{P.~B.} \bibnamefont{Littlewood}},
  \bibinfo{journal}{New Journal of Physics} \textbf{\bibinfo{volume}{14}},
  \bibinfo{pages}{105008} (\bibinfo{year}{2012}).

\bibitem[{\citenamefont{Magnus}(1954)}]{magnus1954}
\bibinfo{author}{\bibfnamefont{W.}~\bibnamefont{Magnus}},
  \bibinfo{journal}{Comm. Pure Appl. Math} \textbf{\bibinfo{volume}{VII}},
  \bibinfo{pages}{649} (\bibinfo{year}{1954}).

\bibitem[{\citenamefont{Knuth}(1998)}]{knuth1998}
\bibinfo{author}{\bibfnamefont{D.~E.} \bibnamefont{Knuth}},
  \emph{\bibinfo{title}{The Art of Computer Programming - Vol 2 - Seminumerical
  Algorithms}} (\bibinfo{publisher}{Addison-Wesley}, \bibinfo{year}{1998}).

\bibitem[{\citenamefont{Nyquist}(1928)}]{nyquist1928}
\bibinfo{author}{\bibfnamefont{H.}~\bibnamefont{Nyquist}},
  \bibinfo{journal}{AIEE Trans.} \textbf{\bibinfo{volume}{47}},
  \bibinfo{pages}{617} (\bibinfo{year}{1928}).

\bibitem[{\citenamefont{Shannon}(1949)}]{shannon1949}
\bibinfo{author}{\bibfnamefont{C.~E.} \bibnamefont{Shannon}},
  \bibinfo{journal}{Proc. Inst. of Radio Eng.} \textbf{\bibinfo{volume}{37}},
  \bibinfo{pages}{10} (\bibinfo{year}{1949}).

\bibitem[{\citenamefont{Donoh}(2006)}]{donoh2006}
\bibinfo{author}{\bibfnamefont{D.}~\bibnamefont{Donoh}}, \bibinfo{journal}{IEEE
  Trans. Inf. Theory} \textbf{\bibinfo{volume}{52}}, \bibinfo{pages}{1289}
  (\bibinfo{year}{2006}).

\bibitem[{ric()}]{rice}
\urlprefix\url{http://dsp.rice.edu/cs}.

\bibitem[{\citenamefont{Needell}(2009)}]{needell2009}
\bibinfo{author}{\bibfnamefont{D.}~\bibnamefont{Needell}}, Ph.D. thesis,
  \bibinfo{school}{UC Davis} (\bibinfo{year}{2009}),
  \urlprefix\url{http://statweb.stanford.edu/~dneedell/papers/dissertation_ss.pdf}.

\bibitem[{\citenamefont{Wright et~al.}(2009)\citenamefont{Wright, Nowak, and
  Figueiredo}}]{wright2009}
\bibinfo{author}{\bibfnamefont{S.~J.} \bibnamefont{Wright}},
  \bibinfo{author}{\bibfnamefont{R.~D.} \bibnamefont{Nowak}}, \bibnamefont{and}
  \bibinfo{author}{\bibfnamefont{M.~A.~T.} \bibnamefont{Figueiredo}},
  \bibinfo{journal}{IEEE Trans. on Signal Process.}
  \textbf{\bibinfo{volume}{57}}, \bibinfo{pages}{2479} (\bibinfo{year}{2009}).

\bibitem[{\citenamefont{Ward}(2009)}]{ward2009}
\bibinfo{author}{\bibfnamefont{R.}~\bibnamefont{Ward}}, \bibinfo{journal}{IEEE
  Trans. Inform. Theory} \textbf{\bibinfo{volume}{55}}, \bibinfo{pages}{5773}
  (\bibinfo{year}{2009}).

\end{thebibliography}
\end{document}